\documentclass{article}
\usepackage{arxiv}
\usepackage[utf8]{inputenc} 
\usepackage[T1]{fontenc}    
\usepackage{indentfirst}
\usepackage{hyperref}       
\usepackage{url}            
\usepackage{booktabs}       
\usepackage{amsfonts}       
\usepackage{nicefrac}       
\usepackage{microtype}      
\usepackage{lipsum}		
\usepackage{graphicx}
\usepackage{natbib}
\usepackage{doi}
\usepackage{natbib,bm}
 \bibpunct[, ]{(}{)}{,}{a}{}{,}%
\usepackage{float}
\usepackage{amsmath,amssymb,amsthm}
\usepackage{enumitem} 
\usepackage{amsmath,amssymb}
\usepackage{bm}
\usepackage{graphicx}
\usepackage{subcaption}
\usepackage{amsmath}
\usepackage{amsfonts}
\usepackage{tikz}
\usetikzlibrary{arrows.meta}
\usepackage[justification=centering]{caption}
\theoremstyle{definition}
\newtheorem{definition}{Definition}
\theoremstyle{Theorem}
\newtheorem{Theorem}{Theorem}
\newtheorem{theorem}{Theorem}[section]
\newtheorem{lemma}[theorem]{Lemma}

\newtheorem{remark}[theorem]{Remark}

\usepackage{tikz}
\usetikzlibrary{shapes.geometric, arrows, positioning}
\newcommand{\R}{\mathbb{R}}
\newcommand{\sm}{\sigma_{\rm min}}
\newcommand{\bQ}{\bm{Q}}
\newcommand{\bq}{\bm{q}}
\newcommand{\bqs}{\bm{q},\sigma}

\title{Privacy Data Pricing: A Stackelberg Game Approach}

\date{December 20, 2025} 					

\author{{\hspace{1mm}Lijun~Bo}\\
	School of Mathematics and Statistics\\
    Xidian University, Xi'an, 710126, China.\\
	\texttt{lijunbo@xidian.edu.cn} \\
	\And
	{\hspace{1mm}Weiqiang~Chang}\\
	School of Mathematics and Statistics\\
    Xidian University, Xi'an, 710126, China.\\ 
	\texttt{changweiqiang@stu.xidian.edu.cn}
}

\renewcommand{\shorttitle}{Privacy Data Pricing}

\hypersetup{
pdftitle={A template for the arxiv style},
pdfsubject={q-bio.NC, q-bio.QM},
pdfauthor={David S.~Hippocampus, Elias D.~Striatum},
pdfkeywords={First keyword, Second keyword, More},
}

\begin{document}
\maketitle

\begin{abstract}
	Data markets are emerging as key mechanisms for trading personal and organizational data. Traditional data pricing studies---such as query-based or arbitrage-free pricing models---mainly emphasize price consistency and profit maximization but often neglect privacy constraints and strategic interactions. The widespread adoption of differential privacy (DP) introduces a fundamental privacy--utility trade-off: noise protects individuals' privacy but reduces data accuracy and market value. This paper develops a Stackelberg game framework for pricing DP data, where the market maker (leader) sets the price function and the data buyer (follower) selects the optimal query precision under DP constraints. We derive the equilibrium strategies for both parties under a balanced pricing function where the pricing decision variable enters linearly into the original pricing model. We obtain closed-form solutions for the optimal variance and pricing level, and determine the boundary conditions for market participation. Furthermore, we extend the analysis to Stackelberg games involving nonlinear power pricing functions. The model bridges DP and economic mechanism design, offering a unified foundation for incentive-compatible and privacy-conscious data pricing in data markets.
\end{abstract}

\section{Introduction}\label{sec:intro}

Data has rapidly become a critical economic asset, giving rise to \textit{data markets} where individuals and organizations buy and sell data. Over the past decade, a number of online market maker have emerged to facilitate such exchanges---for example, services like Xignite for financial data, Gnip for social media feeds, PatientsLikeMe for health statistics, and AggData for aggregated web data. Major marketplaces, such as, the Windows Azure DataMarket and Infochimps have offered thousands of datasets across domains to meet the growing demand. This proliferation of data markets highlights the need for principled data pricing mechanisms. Unlike traditional goods, data products can be queried or copy in many ways, so traditional pricing models  may fail to capture their value (see e.g., \citet{curry2016}). Previous studies have examined data pricing from multiple perspectives, including comprehensive surveys of existing models and issues (see \citet{fricker2017} and \citet{pei2020}), the design of arbitrage-free query-based pricing mechanisms (see \citet{Koutris2015}), and revenue-maximizing strategies for competitive data markets (see \citet{Kushal2012}), collectively laying the groundwork for emerging data marketplaces.

At the same time, the rise of data markets has been paralleled by growing privacy concerns. Much of the traded data is personal or sensitive---from individual online behavior to health and financial records---raising the question of how to protect individuals' privacy when their data is sold or analyzed. Incidents of data misuse and stricter regulations  reflect an increasing insistence that privacy preservation be built into data transactions (see e.g., \citet{Federal} and \citet{Zheng2021}). As awareness has grown, so too has the expectation that individuals be compensated for the privacy risks they incur. In fact, one paradigm shift has been to treat privacy itself as a commodity: individuals sell a certain amount of privacy in exchange for compensation, and the price of data should reflect the privacy loss to those individuals. This perspective demands an approach to pricing that directly accounts for privacy. A promising framework to achieve this is differential privacy (DP), which has become a  standard for quantifying privacy loss (see e.g., \citet{dwork2006}). Differential privacy provides a rigorous definition of privacy protection by guaranteeing that the outcome of a query  is essentially equally likely whether or not any single individual’s data is included. It achieves this by injecting calibrated random noise into query answers, thereby limiting the information leaked about any one individual. Crucially, differential privacy comes with the concept of a privacy budget $\varepsilon$ that quantifies the allowed privacy loss: answering more queries or with higher accuracy (less noise) consumes more budget. This tool has enabled researchers to begin integrating privacy into pricing models. 

Existing literature has explored the integration of differential privacy (DP) with pricing mechanisms across multiple domains.
On one hand, a series of studies have examined the fundamental tension between  privacy and data utility (see e.g., \cite{DaiIEEE2025}, \citet{w2025} and \citet{z2023}). On the other hand, a number of studies have directly incorporated differential privacy (DP) into data pricing frameworks. This body of work encompasses two main branches, both theoretical and applied. At the theoretical level, seminal works such as Roth address the procurement of private data through auction mechanisms (see e.g., \cite{roth2012}). Furthermore, Aperjis et al.  design a market framework that enables individuals to sell unbiased data based on their personal privacy preferences(see e.g. \citet{Aperjis2012}). On the applied front, Lei et al.  explore how firms can utilize privacy-protected data for personalized pricing within revenue management (see e.g., \citet{lei2024}). Other work focuses on personalized pricing, where a seller uses private consumer data to optimize prices for goods or services (see e.g., \citet{chen2023}). Notably, a representative example is the work of Li et al. They advance a theoretical framework for pricing private data by marrying differential privacy with data pricing principles (see e.g., \citet{Li2014}). In their framework, noisy aggregate query answers are priced as a function of their accuracy, and the revenue is apportioned to compensate each data owner for their loss of privacy.

Incorporating differential privacy into data pricing, however, raises new challenges for pricing design. First, any privacy-preserving mechanism limits data accuracy---the noise that protects privacy also diminishes the utility of the data to buyers (see e.g. \citet{alvim2011} and \citet{kim2021}). This creates a fundamental trade-off between privacy and the value of data.  Pricing schemes must carefully adjust for this trade-off. Second, a result from privacy theory is that one cannot release an unbounded number of accurate answers without violating privacy; in practice there is a strict limit on how many queries can be answered. This means data markets cannot simply sell unlimited query access, they must price queries in a way that also manages cumulative privacy loss. \citet{Li2014} addressed this by pricing noisy query answers in an arbitrage-free manner, ensuring that prices increase with accuracy  to reflect higher privacy cost. Each individual’s data contributes a share of the query’s price as a micro-payment compensating them for their privacy loss (see e.g.,  \citet{Ghosh2011}). A challenge here is defining a price function that is privacy-aware yet still arbitrage-free and incentive-compatible (see e.g., \citet{zhang2021}). Differential privacy quantifies privacy loss in a way that can be mapped to monetary value, but  adding noise means query answers are probabilistic; The market maker must assure the buyer that, given the uncertainty in data quality, the buyer can specify the desired query precision.

At present, the existing literature has explored a variety of data pricing approaches from different perspectives, each addressing these issues from its own viewpoint.
 One established line of research concentrates on arbitrage-free pricing for database queries, developing frameworks to ensure price consistency and prevent exploitative bundling (see, e.g., \citet{chen2022}). In parallel, another significant strand employs auction-based mechanisms, designing truthful markets to procure private data from multiple owners or to allocate data access efficiently (see, e.g., \citet{he2024}). Furthermore, a complementary approach focuses on pricing personal data via statistical samples, where buyers purchase access to randomized data subsets rather than full datasets, balancing cost with estimation accuracy (see, e.g., \citet{Gkatzelis2015}).
However, neither classical privacy pricing models nor optimization models systematically account for the dynamic strategic interactions among participants in data markets, which constitutes a primary motivation for introducing a game theoretic perspective in this work.

To address this  gap in modeling strategic interactions, a growing body of research has introduced game-theoretic frameworks to explicitly capture the dynamics of data markets. Such as by integrating phase transition theory  from statistical physics into game theory, established a universal quantitative framework for valuing information in games of chance (see e.g., \citet{xiao2020} and  \citet{Gamberi2024});
Study how pricing and endogenous information acquisition interact within network structures (see e.g., \citet{xiong2025} and \citet{xu2020} ), while the other experimentally analyzes consumer strategic responses to personalized pricing and the consequent demand for privacy (see e.g., \citet{chenz2020} and  \citet{bo2024}). But, it is important to note that the analytical focus of these works has generally not extended to systematically modeling personal data privacy leakage or incorporating privacy-preserving constraints like differential privacy. 
The contributions of this paper are twofold. First, we propose an integrated framework for privacy-constrained data markets that unifies differential privacy (DP) mechanisms with Stackelberg game theory. This model innovatively treats both the market maker’s pricing strategy and the buyer’s query selection as endogenous strategic decisions within a single analytical structure—a formulation not previously established in existing literature. Second, by analyzing the interplay between query value intensity and a derived critical threshold within this game-theoretic framework, we identify distinct regimes of data trading behavior. For each regime, we derive corresponding equilibrium Stackelberg pricing strategies in closed-form expressions. Key findings include: (i) heightened privacy concerns may push up prices, but excessively high prices may cause the market maker's overall revenue to decline; (ii) profit-maximizing conditions exist wherein market makers can optimize offerings by adjusting privacy parameters; and (iii) the proposed mechanism remains arbitrage-free against strategic query manipulation, maintaining system integrity even under rational user behavior. 

The remainder of this paper is organized as follows: Section \ref{sec:model} establishes the foundational architecture of our privacy-preserving data pricing framework and formulates the optimal pricing problem as a bilevel Stackelberg game. We formally characterize the strategic interactions between market makers and data consumers. Building upon this model, Section \ref{Game frame} derives closed-form equilibrium pricing strategies for linear original pricing functions. Through rigorous analysis of the interplay between query value intensity and a critical threshold parameter, we demonstrate how market dynamics partition into distinct behavioral regimes, each governed by unique equilibrium conditions. Section \ref{sec:nonlineapricingfcn} generalizes these findings to nonlinear power pricing structures. This extension reveals richer strategic behaviors while maintaining the core theoretical insights established in the linear case. Finally, Section \ref{sec:conclu} concludes the paper.

\section{The Model and Problem Formulation}\label{sec:model}

\begin{figure}[htbp]
\centering%
\begin{tikzpicture}[x=1cm,y=1cm,>=Stealth,thick,font=\scriptsize]

\node[draw,rounded corners,minimum width=1.8cm,minimum height=0.6cm,
      align=center] (u1) at (-2,4) {User 1};
\node[draw,rounded corners,minimum width=1.8cm,minimum height=0.6cm,
      align=center] (u2) at (-2,3) {User 2};
\node[draw,rounded corners,minimum width=1.8cm,minimum height=0.6cm,
      align=center] (u3) at (-2,2) {User 3};
\node[draw,rounded corners,minimum width=1.8cm,minimum height=0.6cm,
      align=center] (u4) at (-2,1) {User 4};
\node[draw,rounded corners,minimum width=1.8cm,minimum height=0.6cm,
      align=center] (u5) at (-2,0) {User 5};

\node[anchor=west] at (-0.6,4) {$x_1$};
\node[anchor=west] at (-0.6,3) {$x_2$};
\node[anchor=west] at (-0.6,2) {$x_3$};
\node[anchor=west] at (-0.6,1) {$x_4$};
\node[anchor=west] at (-0.6,0) {$x_5$};
\draw[dashed,-] (-0.45,-0.7) -- (-0.45,-1.5);
\draw[dashed,-] (-1,4) -- (-0.6,4);
\draw[dashed,-] (-1,3) -- (-0.6,3);
\draw[dashed,-] (-1,2) -- (-0.6,2);
\draw[dashed,-] (-1,1) -- (-0.6,1);
\draw[dashed,-] (-1,0) -- (-0.6,0);

\draw[dashed,->] (3.4,4) --node[midway, above] {$\mu_1(\bQ)$} (0,4);
\draw[dashed,->] (3.4,3) --node[midway, above] {$\mu_2(\bQ)$} (0,3);
\draw[dashed,->] (3.4,2) --node[midway, above] {$\mu_3(\bQ)$} (0,2);
\draw[dashed,->] (3.4,1) --node[midway, above] {$\mu_4(\bQ)$} (0,1);
\draw[dashed,->] (3.4,0) --node[midway, above] {$\mu_5(\bQ)$} (0,0);

\node[draw,minimum width=1.8cm,minimum height=5cm,align=center]
       at (4.5,2) {Market Maker\\ \\ \\ \\ \textbf{Leader}};

\draw[dashed,-] (4.5,-0.7) -- (4.5,-1.5);

\draw[dashed,->] (3.5,-1) -- (4.5,-1);

\draw[dashed,->] (1.0,-1) -- (-0.45,-1);

\node[draw,minimum width=1.8cm,minimum height=5cm,align=center]
      (by) at (10,2) {Buyer\\ \\ \\ \\ \textbf{Follower}};

\draw[dashed,-] (10,-0.7) -- (10,-1.5);

\draw[dashed,->] (8.9,-1) -- (10,-1);

\draw[dashed,->] (5.7,-1) -- (4.5,-1);

\draw[dashed,->] (9,4) --node[midway, above] {A1. Linear query: }
node[midway, below]{$\bQ=(\bq,\sigma)$} (5.6,4);
\draw[dashed,->] (5.6,2.7) --node[midway, above] {A2. Pricing function:}
node[midway, below]{$\pi(\bQ)$} (9,2.7);
\draw[dashed,->] (9,1.3) --node[midway, above] {A3. Payment:} 
node[midway, below] {$\pi(\bQ)$} (5.6,1.3);

\draw[dashed,->] (5.6,0) --node[midway, above] {A4. Query Answer:}
node[midway, below] {$\bq({\bm x})+\text{noise}$}(9,0);

\node[anchor=south west] at (-1.3,4.8) {Data Items};
\node at (4.8,5.5)
  {(B) Privacy loss: $\varepsilon_1,\varepsilon_2,\ldots,\varepsilon_{5}$};

\draw[dashed,->] (10,5.5) -- (10,4.5);

\draw[dashed,-] (7,5.5) -- (10,5.5);

\draw[dashed,-] (-1.8,5.5) -- (2.8,5.5);

\draw[dashed,-] (-1.8,4.4) -- (-1.8,5.5);

\node[anchor=north west] at (-3,-0.6)
  {$\bm{x}=(x_1,\ldots,x_{5})$};

\node at (2.2,-1) {(C) Compensation};
\node at (7.3,-1) {(A) Pricing \& Purchase};

\end{tikzpicture}
\caption{The market-maker facilitates interactions between buyers and sellers.}\label{fig:1}
\end{figure}
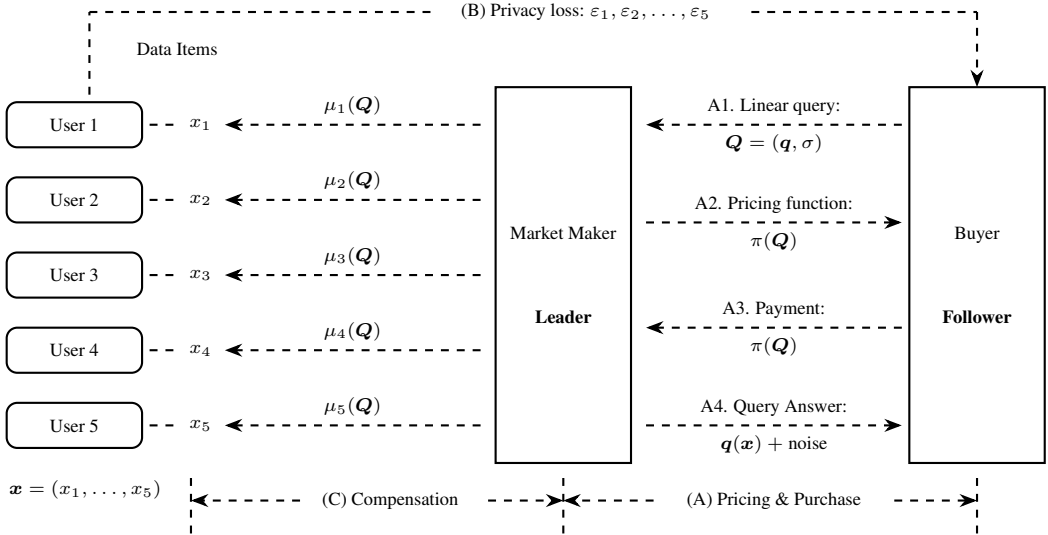

In this section, we describe the basic architecture of the private data pricing framework considered in the paper, illustrated in Figure \ref{fig:1}.

Figure \ref{fig:1} 
depicts the basic architecture of the privacy data market and the roles of the three parties. On the left, multiple data owners each hold a personal data item, and together they form the database used to answer the buyer's query $\bQ=(\bqs)$. In the middle, the market maker acts as an intermediary: it commits to a pricing function $\pi(\bQ)$ and a compensation rule $\mu_i(\bQ)$, adds calibrated noise to $\bq({\bm x})$, collects the buyer's payment, and redistributes micro-payments to data owners. On the right, the buyer submits the query, observes the posted price, decides whether to purchase, and receives the noisy answer $\bq({\bm x})+\text{noise}$. The arrows summarize the flows of payments, privacy loss, and compensation within this Stackelberg game based privacy data market.

\subsection{The Main Actors}

The main actors consist of the following market participants:  

\begin{itemize}
    \item {\it The market maker.} In our setting, a trusted market maker intermediates between the buyer and individual data owners. The intermediary acquires data from owners and sells access via query responses. When a buyer purchases a query, the market maker collects the fee, computes the requested statistic, injects the noise required by the privacy policy, returns the perturbed result to the buyer, and remits individual payments to the data owners. The market maker retains a portion of the transaction price as profit.
    \item {\it The data vendor.} A database is a vector of
real-valued data items $\bm{x}=( x_{1},\ldots, x_{n})\in\mathbb{R}^n$, which contains the complete information of $n$ data owners. The data item $ x_{i}$ represents personal information, owned
by $i$-th individual. For example, consider a table with attributes \textit{age}: entries $x_1$, $x_2$, and $x_3$ represent the information of the first, second, and third data owners, respectively.

\item {\it The buyer with linear queries.} The buyer is a data analyst who wishes to compute some queries over the data. We consider the class of linear aggregation queries over the data items in database $\bm{x}\in\R^n$.
\end{itemize}

\begin{definition}[Linear Query]\label{def:linear-query}
A linear query is a real-valued vector $\bm{q}=(q_1,\ldots,q_n)\in\mathbb{R}^n$. The answer $\bm{q}(\bm{x})$ to this linear query on the database $\bm{x}=(x_1,\ldots,x_n)\in\mathbb{R}^n$ is the vector product 
$\bm{q}(\bm{x})={\bm q}\cdot{\bm x}:=q_1x_1+\cdots+q_nx_n$.
\end{definition}
We assume that the buyer is allowed to issue multiple queries. This means that the buyer can combine information derived from multiple queries to infer answers to other queries not explicitly requested. 

\subsection{Privacy Loss}
Our definition of privacy loss is adapted from DP, which compares the output of
a mechanism with and without the contribution of the data item $x_i\in\mathbb{R}$. Herein, we adopt the Laplace mechanism to achieve differential privacy, whose definition is provided below:

\begin{definition}[Laplacian Mechanism (LM)]\label{def:L-M}
    The LM denoted by $\mathcal{L}$ is the data-independent mechanism defined as, for a given query ${\bm Q}=(\bm{q},\sigma)\in\mathbb{R}^n\times(0,\infty)$ and a database instance $\bm{x}\in\mathbb{R}^n$, the corresponding mechanism answer $\mathcal{L}_{{\bm Q}}(\bm{x})= \bm{q}(\bm{x})+\xi$, where $\xi\sim\operatorname{Lap}(0,\sqrt{\sigma / 2})$ is the noise. Here, $\operatorname{Lap}(a,b)$ represents the Laplace distribution with location parameter $a\in\mathbb{R}$ and scale parameter $b>0$.
\end{definition}

In the query ${\bm Q}=(\bm{q},\sigma)\in\mathbb{R}^n\times(0,\infty)$, $\sigma=\operatorname{Var}[\xi]>0$ is the {\it answers' noise  variance}, which is added under the DP mechanism, captures the accuracy of the released answer: lower $\sigma$ implies higher precision. This reflects the fundamental trade-off between privacy and accuracy. Consider an efficient market consisting of $n$ data owners. In this market, data should be priced at least high enough to compensate the data owner for the incurred privacy loss. We next state a lemma that provides an upper bound on the $i$-th data owner’s privacy loss under the LM in Definition~\ref{def:L-M}.

\begin{lemma}[\citet{Dwork2006}]\label{lem:Dwork}
Let $\mathcal{L}_{\bm Q}$ be the LM corresponding to a query ${\bm Q}=(\bm{q},\sigma)\in\mathbb{R}^n\times(0,\infty)$. Then, the privacy loss of individual $i$, denoted by $\varepsilon_i\left(\mathcal{L}_{{\bm Q}}\right)$, is bounded by
\begin{align}\label{Privacy Loss}
\varepsilon_i\left(\mathcal{L}_{{\bm Q}}\right) \leq \frac{s_i}{\sqrt{\sigma / 2}},\quad \forall i=1,\ldots,n,
\end{align}
where $s_i:=\max_{\bm{x}\in \mathbb{R}^n}|\bm{q}(\bm{x})-\bm{q}(\bm{x}^{(i)})|$ is the personalized sensitivity associated with item $i$. Here,  $\bm x^{(i)}=(x_1,\ldots,x_{i-1},x_i',x_{i+1},\ldots,x_n)\in \mathbb{R}^n$ is the corresponding neighbor-database of ${\bm x}=(x_1,\ldots,x_{i-1},x_i,x_{i+1},\ldots,x_n)\in \mathbb{R}^n$ when they differ only in the $i$-th data item. 
\end{lemma}

We here consider the linear query $\bm{q}(\bm{x})=\bm{q}\cdot\bm{x}$ with ${\bm Q}=(\bm{q},\sigma)\in\mathbb{R}^n\times(0,\infty)$. To this purpose, for the neighbor-database $\bm{x}^{(i)}$ and $\bm{x}$ in Lemma~\ref{lem:Dwork}, let $x_i'=x_i+\gamma_i$ with $\gamma_i\in\mathbb{R}$ being the given change quantity. In this case, the personalized sensitivity associated with item $i$ is given by $s_i=\max_{x^{(i)}\in\mathbb{R}^n}|\bm{q}\cdot \bm{x}-\bm{q}\cdot \bm{x}^{(i)}|\leq\gamma|q_i|$ with $\gamma:=\max_{i=1}^n |\gamma_i|$. As a consequence, the bound \eqref{Privacy Loss} becomes that
    \begin{align}\label{Privacy Loss2}
\varepsilon_i\left(\mathcal{L}_{{\bm Q}}\right) \leq \frac{\gamma|q_i|}{\sqrt{\sigma / 2}},\quad\forall i=1,\ldots,n.
\end{align}

\subsection{Micro-payment }
We describe the third component of the pricing framework in Figure \ref{fig:1}, namely, the micro-payments $\mu_i(\bm{q},\sigma)$ with the linear query $\bm{Q}=(\bm{q},\sigma)\in\mathbb{R}^n\times(0,\infty)$. When answering a buyer’s query $\bm{Q}$ using a mechanism $\mathcal{L}_{\bm{Q}}$, the market maker inevitably leaks some private information about the data owners and must compensate each owner with a micro-payment $\mu_i(\bm{q},\sigma)$ for every data item $x_i$ they own. Micro-payments close the loop in Figure \ref{fig:1}, that is, they must be covered by the buyer’s payment $\pi$ and must depend on the degree of privacy loss $\varepsilon_i$. Then, by Lemma~\ref{lem:Dwork} with \eqref{Privacy Loss2} and Proposition 32 of \citet{Li2014}, the micro-payment function is taken as the following form:
\begin{align}\label{eq:muiQ}
\mu_i(\bm{q},\sigma)=\gamma\frac{c_{i}|q_i|}{\sqrt{\sigma/2}},\quad \forall i=1,\ldots,n,
\end{align}
where $c_i>0$ is the privacy weight of the $i$-th data owner. The micro-payment function \eqref{eq:muiQ} ensures that compensation is proportional to the privacy loss, which depends on the query’s sensitivity to the data owner’s information and the precision of the output. The privacy weight $c_i$ allows for adjustments based on the relative value of the data privacy owned by different individuals.

\subsection{Utility Functions}
We consider a setting with two parties: a market maker and a data buyer. The market maker holds a sensitive dataset and, while protecting the privacy of the $n$ data owners, sells answers to queries over this data to the buyer. The interaction can be viewed as a leader–follower scenario in which the market maker first chooses the privacy level and price for the linear query $\bm{Q}=(\bm{q},\sigma)\in\R^n\times(0,\infty)$, and the buyer then decides whether to purchase the data at that price given its expected accuracy. Now, we outline the key elements of the model as follows:

\begin{itemize}

\item \textit{The original pricing function.} We follow the pricing framework as in \citet{Li2014} to introduce the original pricing function which is given by, for a query $\bQ=(\bqs)\in\R^n\times(0,\infty)$, 
\begin{align}\label{eq:originalpricingfcn}
 \pi_0(\bqs;k):=k\frac{f(\bq)^2}{\sigma},   
\end{align}
where $k>0$ is the pricing decision variable in our optimal pricing framework. The pricing decision variable $k>0$ plays a critical role in modulating price sensitivity relative to query characteristics. Specifically, it enables dynamic adjustment of the pricing function based on the norm of the query vector $\bq$, which jointly influences both the accuracy guarantees and computational costs associated with query resolution (c.f. Example 3.17 in \citet{Li2014}). Here, $f:\R^n\to[0,\infty)$ is a {\it semi-norm}, i.e., it satisfies (i) $f(\alpha\bq)=|\alpha|f(\bq)$ for any $\alpha\in\R$ and $\bq\in\R^n$; and (ii) $f(\bq_1+\bq_2)\leq f(\bq_1)+f(\bq_2)$ for any $\bq_1,\bq_2\in\R^n$. For example, the $L^2$-norm $f(\bq)=\|\bm{q}\|_2:=\sqrt{\sum_{j=1}^nq_j^2}$ of a query vector $\bm{q}\in\R^n$ is a norm, and hence is a semi-norm. 

\item \textit{The balanced pricing function $\pi({\bm q},\sigma;k)$.} The price charged by the market maker for answering a query $\bm{Q}=(\bm{q},\sigma)\in\mathbb{R}^n\times(0,\infty)$. The pricing decision variable $k \geq 0$ is defined in the original pricing function $\pi_0$. Herein, we consider so-called {\it balanced pricing function}, which is described as follows:
\begin{align}\label{pricing function}
\pi(\bm{q},\sigma;k)=\max\left\{\pi_0(\bqs;k),\ \sum_{i=1}^{n}\mu_i(\bm q,\sigma)\right\}.
\end{align}
The balanced pricing mechanism entails that the market maker establishes two contractual obligations: (i) it commits to fulfilling buyer queries at a predetermined price $\pi_0$; and (ii) it agrees to redistribute micro-payments $\mu_i$ for $i=1,\ldots,n$ to data owners as compensation for privacy losses incurred when their data is processed in response to buyer queries. Therefore, the pricing function $\pi(\bq,\sigma;k)$ defined in \eqref{pricing function} can ensure that the market maker has non-negative profits.
\end{itemize}

We next introduce the arbitrage-freeness of the pricing function $\pi(\bQ)$ as in \citet{Li2014}. For any $m\geq1$, the query $\bQ=(\bqs)$ is said to be linear answerable from query sequence $\bm{S}=\{\bm Q_j=(\bq_j,\sigma_j)\}_{j=1}^m$, if there exist $\bm{\alpha}=(\alpha_1, \ldots,\alpha_m)\in\R^m$ such that $\sum_{j=1}^m\alpha_j\bq_j=\bq$ and $\sum_{j=1}^m\alpha_j^2\sigma_j\leq\sigma$. Then, the pricing function $\pi(\bm Q)$ is arbitrage-free, if for any linear answerable query sequence $\bm S$ satisfying
\begin{align}\label{arbitrage-free}
\pi(\bm Q)\leq \sum_{j=1}^{m}\pi (\bm Q_{j}).
\end{align}

Then, we have
\begin{lemma}\label{lem:arbitrage-free}
Let $k>0$ be fixed. For the query $\bm{Q}=(\bm{q},\sigma)\in\mathbb{R}^n\times(0,\infty)$, the balanced pricing function $\pi(\bm{Q};k)$ given by \eqref{pricing function} is arbitrage-free.   
\end{lemma}

{\it Proof.}\quad Let $\bm{S}=\{\bm Q_j=(\bq_j,\sigma_j)\}_{j=1}^m$ with  $\bq_j=(q_1^{j},\ldots,q_n^j)\in\R^n$ be an arbitrary linear answerable query sequence, i.e., there exist $\bm{\alpha}=(\alpha_1, \ldots,\alpha_m)\in\R^m$ such that $\sum_{j=1}^m\alpha_j\bq_j=\bq$ and $\sum_{j=1}^m\alpha_j^2\sigma_j\leq\sigma$. Then, in view of \eqref{pricing function} and \eqref{eq:muiQ}, we have
\begin{align}\label{eq:CS1}
  \sum_{j=1}^{m}\pi (\bm Q_{j};k) &= \sum_{j=1}^m \max\left\{\frac{kf(\bm q_j)^2}{\sigma_j},\ \sum_{i=1}^{n}\mu_i(\bq_j,\sigma_j)\right\}=\sum_{j=1}^m \max\left\{\frac{kf(\bm q_j)^2}{\sigma_j},\ \sum_{i=1}^{n}c_{i}\frac{\gamma|q_i^j|}{\sqrt{\sigma_j/2}}\right\}\nonumber\\
  &=\max\left\{k\sum_{j=1}^m\frac{f(\bm q_j)^2}{\sigma_j},\ \sum_{i=1}^{n}c_{i} \gamma \left(\sum_{j=1}^m\frac{|q_i^j|}{\sqrt{\sigma_j/2}}\right)\right\}.
\end{align}
It follows from Cauchy-Schwarz inequality that
\begin{align}\label{eq:ine1}
f(\bq)^2&=f\left(\sum_{j=1}^m(\alpha_j\bq_j)\right)^2\leq \left(\sum_{j=1}^mf(\alpha_j\bq_j)\right)^2=\left(\sum_{j=1}^m |\alpha_j|\sqrt{\sigma_j}\frac{f(\bq_j)}{\sqrt{\sigma_j}}\right)^2\leq \left(\sum_{j=1}^m\alpha_j^2\sigma_j\right)\left(\sum_{j=1}^m\frac{f(\bq_j)^2}{\sigma_j}\right)\nonumber\\
  &\leq\sigma \left(\sum_{j=1}^m\frac{f(\bq_j)^2}{\sigma_j}\right)\implies \sum_{j=1}^m\frac{f(\bq_j)^2}{\sigma_j}\geq\frac{f(\bq)^2}{\sigma}.
\end{align}
On the other hand, we have
\begin{align}\label{eq:ine2}
    \sum_{i=1}^{n}c_{i}\gamma\frac{|q_i|}{\sqrt{\sigma/2}}&\leq  \sum_{i=1}^{n}c_{i}\gamma\frac{\left|\sum_{j=1}^m\alpha_jq_i^j\right|}{\sqrt{\sum_{k=1}^m\alpha_k^2\sigma_k/2}}\leq \sum_{i=1}^{n}c_{i}\gamma\left(\sum_{j=1}^m\frac{\left|\alpha_j\right|\left|q_i^j\right|}{\sqrt{\sum_{k=1}^m\alpha_k^2\sigma_k/2}}\right)\nonumber\\
    &\leq \sum_{i=1}^{n}c_{i}\gamma\left(\sum_{j=1}^m\frac{\left|\alpha_j\right|\left|q_i^j\right|}{|\alpha_j|\sqrt{\sigma_j/2}}\right)=\sum_{i=1}^{n}c_{i}\gamma\left(\sum_{j=1}^m\frac{|q_i^j|}{\sqrt{\sigma_j/2}}\right).
\end{align}
By applying \eqref{eq:ine1} and \eqref{eq:ine2}, we have from \eqref{pricing function} and \eqref{eq:CS1} that
\begin{align*}
  \sum_{j=1}^{m}\pi (\bm Q_{j};k) &=\max\left\{k\sum_{j=1}^m\frac{f(\bm q_j)^{2}}{\sigma_j},\ \sum_{i=1}^{n}c_{i}\gamma\left(\sum_{j=1}^m\frac{|q_i^j|}{\sqrt{\sigma_j/2}}\right)\right\}\nonumber\\
  &\geq\max\left\{\frac{kf(\bq)^{2}}{\sigma},\ \sum_{i=1}^{n}c_{i}\gamma\frac{|q_i|}{\sqrt{\sigma/2}}\right\}=\pi(\bQ;k). 
\end{align*}
This yields from \eqref{arbitrage-free} that the balanced pricing function $\pi(\bm{Q};k)$ given by \eqref{pricing function} is arbitrage-free. \hfill$\Box$

\begin{itemize}
\item {\it The buyer’s valuation $v(\bqs)$.} The buyer’s valuation for the query $\bm{Q}=(\bm{q},\sigma)\in\mathbb{R}^n\times(0,\infty)$. Following \citet{BergemannMorris2019}, the buyer’s gross informational value from the query $\bm{Q}=(\bqs)\in\R^n\times(0,\infty)$ is given by
\begin{align} \label{eq:buyer_value}
  V({\bm q},\sigma)=\frac{A({\bm q})}{\sqrt{\sigma}},
\end{align}
where $A({\bm q})>0$ is the query’s value intensity, a query-specific coefficient that measures the importance of the query. The classical instant is $A(\bm q)=\ln(1+\|\bm q\|_{0})$, where $\|\bm q\|_{0}$ counts the number of data items engaged by the linear query, which is defined as the number of non-zero elements in the vector $\bq\in\R^n$. Since the logarithmic function is concave, $A(\bm q)$ exhibits diminishing marginal utility as additional data are incorporated. This implies that the buyer’s valuation increases with accuracy while exhibiting diminishing marginal returns.
\end{itemize}

We now proceed to introduce the utility functions of the buyer and the market maker, respectively. The buyer’s utility for purchasing the query $\bm{Q}=(\bm{q},\sigma)\in\R^n\times(0,\infty)$ is defined by the difference between their valuation and the price set by the market maker for this query:
\begin{align}\label{eq:Phi}
        \Phi({\bm q},\sigma;k):=v({\bm q},\sigma)-\pi({\bm q},\sigma;k)=\frac{A({\bm q})}{\sqrt{\sigma}}-\max\left\{\frac{kf(\bq)^{2}}{\sigma},\ \sum_{i=1}^{n}\gamma\frac{c_{i}|q_i|}{\sqrt{\sigma/2}}\right\}.
\end{align}
On the other hand, the market maker is a platform that intermediates data owners and buyers and commits to a pricing schedule for noisy query answers and  a compensation rule for owners' privacy losses. The market maker collects revenue from the buyer according to the price function $\pi(\bm q,\sigma;k)$ and disburses micro-compensations $\mu_i(\bm q,\sigma)$ to each owner $i\in\{1,\dots,n\}$. Then, the market maker’s  utility equals revenue minus the aggregate compensation cost:
\begin{align} \label{eq:mm_objective}
\Psi(\bm{q},\sigma;k):=\pi(\bqs;k)-\sum_{i=1}^{n}\mu_i(\bqs).
\end{align}
By utilizing \eqref{eq:muiQ} and \eqref{pricing function}, we obtain 
\begin{align}\label{eq:mm_utility}
\Psi(\bqs;k)=\max\left\{\frac{kf(\bm{q})^{2}}{\sigma},~\sum_{i=1}^{n}c_{i}\gamma\frac{|q_i|}{\sqrt{\sigma/2}}\right\}-\sum_{i=1}^{n}c_{i}\gamma\frac{|q_i|}{\sqrt{\sigma/2}}=\left(\frac{kf(\bm{q})^{2}}{\sigma}-\sum_{i=1}^{n}\gamma\frac{c_{i}|q_i|}{\sqrt{\sigma/2}}\right)^+,
\end{align}
where $x^+:=\max\{x,0\}$ is the positive part of the real-number $x\in\R$. In particular, we consider the transaction is indexed by the query ${\bm Q}=({\bm q},\sigma)\in\mathbb{R}^n \times[\sigma_{\rm min}(\bq),+\infty)$, where $\sigma\geq \sm(\bq)>0$ is the announced variance of the additive noise used to answer the query; while $\sm(\bq)$ is the market maker’s minimum tradable variance which may vary with respect to the query $\bq$. 

\section{Stackelberg Game Problem}\label{Game frame}

In this section, we study how to jointly determine the optimal pricing and level of privacy for the  market maker  and buyer, aiming to maximize their respective utilities, formulated as a two-stage Stackelberg game problem.

The two-stage Stackelberg game problem is described as follows:
\begin{itemize}
\item {\bf Stage I}~({\it Leader--Market Maker}). The market maker commits to a mechanism characterized by a price function $\pi(\bqs;k)$ and a compensation rule $\{\mu_i(\bqs)\}_{i=1}^n$, where $k>0$ is the mechanism parameter. The market maker’s objective is to choose an optimal pricing parameter $k^*(\bqs)$ so to maximize its own utility $\Psi(\bqs;k)$ over $k\in(0,\infty)$ when the query $\bQ=(\bqs)$ is given:
\begin{align}\label{eq:Stage1}
k^*(\bqs):= \arg\!\max_{k>0}\Psi(\bqs;k)=\arg\!\max_{k>0}\left[\pi(\bqs;k)-\sum_{i=1}^{n}\mu_i(\bqs)\right].
\end{align}

\item {\bf Stage II}~({\it Follower--Buyer}). Given $k^*(\bqs)$ provided by \eqref{eq:Stage1}. The buyer observes the announced price function and selects an optimal precision level (noise variance) $\sigma^*(\bq)$ to maximize its own utility $\Phi(\bqs;k^*(\bqs))$ over $\sigma\geq\sm(\bq)$, i.e.,
\begin{align}\label{eq:Stage2}
\sigma^*(\bq):=\arg\!\!\max_{\sigma\geq\sm(\bq)}\Phi(\bqs;k^*(\bqs))=\arg\!\!\max_{\sigma\geq\sm(\bq)}
\left[V(\bqs)-\pi(\bqs;k^*(\bqs))\right].
    \end{align} 
\end{itemize}
We call $(k^*(\bq),\sigma^*(\bq))=(k^*(\bq,\sigma^*(\bq)),\sigma^*(\bq))$ the equilibrium strategy of the two-stage  Stackelberg game \eqref{eq:Stage1} and \eqref{eq:Stage2}. 

In  Stackelberg games, the leader’s strategy is determined first, as it is based on the anticipation of the follower’s optimal response. This sequential decision-making process arises from the fact that the leader’s choice influences the follower’s behavior, and the follower’s reaction must be considered when optimizing the leader's strategy. Therefore, we apply the method of backward induction in solving this game problem \eqref{eq:Stage1} and \eqref{eq:Stage2}:
\begin{Theorem}\label{The1}
Let the buyer’s utility $\Phi(\bqs;k)$ and the market-maker's utility $\Psi(\bqs;k)$ be respectively given by \eqref{eq:Phi} and \eqref{eq:mm_utility}. Introduce the threshold of the query’s value intensity given by
\begin{align}\label{eq:Gammathreshold}
    \Gamma(\bq):= \sqrt{2}\gamma\sum_{i=1}^{n}c_i|q_i|,
\end{align}
where $\mu_i(\bqs)$ is the micro-payment function of the i-th data item, which is given by \eqref{eq:muiQ}. Then, the  Stackelberg equilibrium strategy $(k^*(\bq),\sigma^*(\bq))=(k^*(\bq,\sigma^*(\bq)),\sigma^*(\bq))$ is given as follows:
\begin{itemize}
    \item[(i)] if  $A(\bm q)\geq 2\Gamma(\bq)$, the market maker’s optimal decision variable and the buyer's variance strategy are given by
    \begin{align}\label{eq:optimalks}
     \left(k^*(\bq),\sigma^*(\bq)\right)=\left(\frac{\sqrt{\sm(\bq)}}{2}\frac{A(\bm q)}{f(\bq)^2},~\sm(\bq)\right).  
    \end{align}

    \item[(ii)] if $\Gamma(\bq)<A(\bm q)<2\Gamma(\bq)$, the utility of market maker becomes that $\Psi(\bm q,\sigma;k) \equiv0$. This implies that any $k>0$ is an optimal decision variable for the market maker. The buyer's optimal variance strategy is given by 
    \begin{align*}
      \sigma^*(\bq,k)=\left(\frac{k}{\Gamma(\bq)}\right)^2f(\bq)^4,\quad \forall k>0.  
    \end{align*}
    
    \item[(iii)] if $A(\bm q) \leq \Gamma(\bq)$,  the utility of the buyer $\Phi(\bm q,\sigma;k) \leq 0$ and $(0,\infty)\ni\sigma\to\Phi(\bqs;k)$ is strictly increasing, and hence the optimal variance is $\sigma^*(\bq)=+\infty$ and the optimal decision variable is any $k>0$. In this case, the buyer does not purchase any data.
\end{itemize}
\end{Theorem}

Economically, Theorem 1 shows that, when $A(\bq)\geq 2\Gamma(\bq)$, i.e., the the query’s value intensity is larger than twice the critical threshold, the buyer always prefers more accurate data, but higher accuracy increases the price and the privacy compensation. In this case, the buyer’s valuation is high enough that, even at the lower bound $\sigma_{\rm min}(\bq)$, the buyer’s utility remains strictly positive. Anticipating this, the market maker chooses $k^*(\bq)$ to make the buyer just indifferent between further increasing $\sigma$ and staying at $\sigma_{\rm min}(\bq)$, while at the same time ensuring that the revenue exceeds the total micro–payments. Thus, $A(\bq)\geq 2\Gamma(\bq)$ describes a profitable trading region where high–value queries support positive profit for the platform.

 When $\Gamma(\bq)<A(\bq)<2\Gamma(\bq)$ i.e., the the query’s value intensity is between the critical threshold and twice the critical threshold, the query’s value is large enough for the buyer to cover the privacy costs, but not large enough for the market maker to earn strictly positive profit once the buyer optimally chooses the precision level. For given 
$k>0$, the buyer chooses $\sigma$ to maximize her net utility. The optimal choice is exactly the threshold $\sigma_{\rm th}(\bq)$ at which the two branches of the balanced pricing function coincide. Any attempt by the market maker to increase  $k$ in order to charge more induces the buyer to adjust $\sigma$ in such a way that the platform again just breaks even. Consequently, the market maker is indifferent over all
$k>0$. Thus, the condition $\Gamma(\bq)<A(\bq)<2\Gamma(\bq)$ describes a break–even region: trade occurs and the buyer benefits, but the market maker can at best operate at zero profit.

 When $A(\bq)\leq\Gamma(\bq)$, i.e., the the query’s value intensity is lower than the critical threshold, it implies that the value intensity of the query $A(\bq)$ is too small relative to the aggregate privacy cost, the buyer cannot obtain non–negative utility from any finite precision level. Then,  $A(\bq)\leq\Gamma(\bq)$  marks a no–trade region: the total value that the buyer can extract from the query is insufficient even to compensate the privacy losses of the data owners, so any transaction would leave the buyer worse off.
\begin{figure}[ht]
    \centering
    \includegraphics[width=0.65\linewidth]{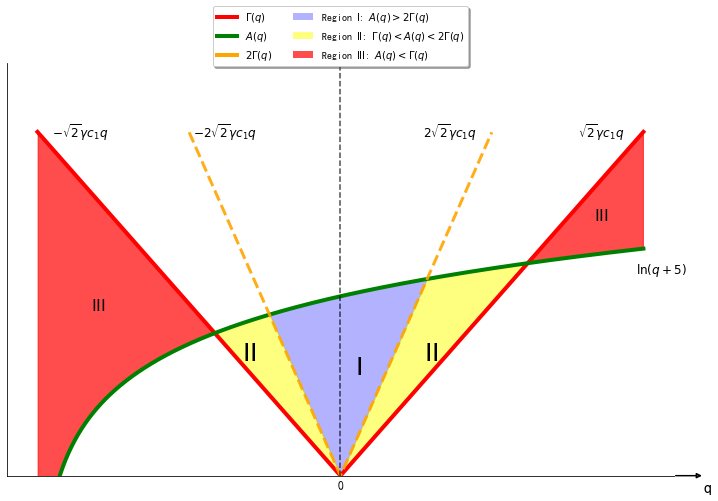}
    \caption{The tradable regions with number of items $n=1$ and the query intensity $A(q)=\ln{(q+5)}$ for $q>-5$.}
    \label{tradable}
\end{figure}
Figure \ref{tradable} illustrates the three distinct market trading regions for a single data item ($n=1$), plotted against the query vector $q$ with intensity function $A(q) = \ln(q+5)$ for $q>-5$. Crucially, the trading behavior is governed by a privacy-cost threshold $\Gamma(q)=\sqrt{2}\gamma c_1|q|$ (thus, the threshold mapping $q \mapsto\Gamma(q)$ exhibits two straight-lines with respective slopes $\sqrt{2}\gamma c_1 > 0$ and $-\sqrt{2}\gamma c_1 < 0$), which divides the domain into three regimes differentiated by color: the profitable trading region blue (see Region I), where $A(q) \geq 2\Gamma(q)=2\sqrt{2}\gamma c_1|q|$ (thus, the threshold mapping $q \mapsto 2\Gamma(q)$ exhibits two straight-lines with respective slopes $2\sqrt{2}\gamma c_1 > 0$ and $-2\sqrt{2}\gamma c_1 < 0$), and the market attains a positive-profit equilibrium; the break-even region yellow, where $\Gamma(q) < A(q) < 2\Gamma(q)$, and transactions occur but yield zero profit to the platform; and the no-trade region red, where $A(q) \leq \Gamma(q)$, and the query value is insufficient to cover privacy costs, resulting in no purchase. The figure thus highlights how the threshold $q\mapsto\Gamma(q)$ critically determines whether trade is feasible, profitable, or unsustainable, visually underscoring the boundary conditions that drive market outcomes under privacy constraints.
\begin{figure}[H]
    \centering
\includegraphics[width=0.6\linewidth]{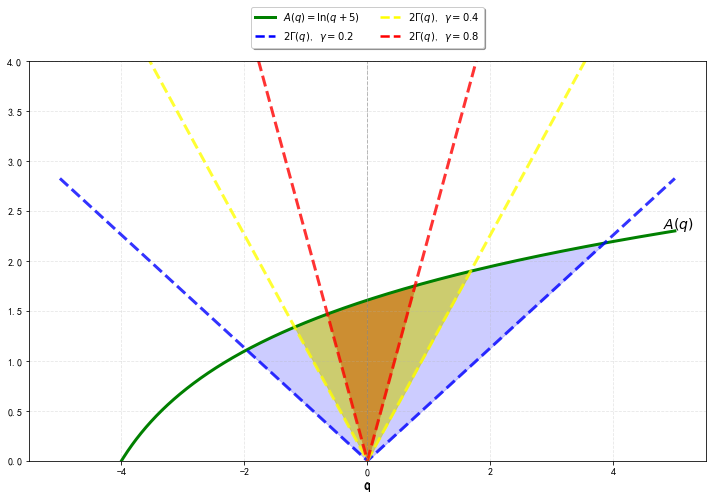}
    \caption{Tradable region boundaries for different values for the privacy cost parameter $\gamma$.}
    \label{gammachange}
\end{figure}

Figure \ref{gammachange} shows how the privacy cost parameter $\gamma$ affects the profitable trading region of the market, defined by the condition $A(q) > 2\Gamma(q)$. For a given $\gamma$, this region consists of the part of the $A(q)$ curve that lies above the corresponding $2\Gamma(q)$ line. Visually, the profitable region shrinks as $\gamma$ grows. This pattern highlights the main role of $\gamma$: a higher $\gamma$ (stricter privacy protection) raises the bar for profitability, so that only queries with high enough value $A(q)$ are worth trading. On the other hand, a lower $\gamma$ lowers this bar, allowing more queries to be traded profitably.

We are now in a position to prove Theorem~\ref{The1}.

\noindent{\it Proof of Theorem~\ref{The1}.}\quad We first solve the above {\bf Stage II} when the decision variable $k>0$ is fixed. To this purpose, recall the pricing function $\pi(\bm{q},\sigma;k)$ given by \eqref{pricing function}. Then, by \eqref{eq:Gammathreshold}, we have
\begin{align}\label{eq:pricingfcnpi}
\pi(\bm{q},\sigma;k)&=\max\left\{\frac{kf(\bm q)^{2}}{\sigma},\ \frac{\Gamma(\bq)}{\sqrt{\sigma}}\right\}= \frac{kf(\bq)^{2}}{\sigma}{\bf1}_{\left\{\frac{kf(\bq)^{2}}{\sigma}\geq\frac{\Gamma}{\sqrt{\sigma}}\right\}} +  \frac{\Gamma(\bq)}{\sqrt{\sigma}}{\bf1}_{\left\{\frac{kf(\bq)^{2}}{\sigma}\leq\frac{\Gamma}{\sqrt{\sigma}}\right\}}.  
\end{align}
Note that, for any $\sigma>0$, we have
\begin{align}\label{eq:prichange}
\frac{kf(\bq)^{2}}{\sigma}\geq\frac{\Gamma(\bq)}{\sqrt{\sigma}}\iff \sigma \leq \sigma_{\rm th}(\bq;k):=\left(\frac{k}{\Gamma(\bq)}\right)^2f(\bq)^4.    
\end{align}
Consequently, it follows from \eqref{eq:pricingfcnpi} that
\begin{align}\label{eq:pricingfcnpi2}
\pi(\bm{q},\sigma;k)&=\max\left\{\frac{kf(\bq)^{2}}{\sigma},\ \frac{\Gamma(\bq)}{\sqrt{\sigma}}\right\}= \frac{kf(\bq)^{2}}{\sigma}{\bf1}_{\left\{\sigma\leq\sigma_{\rm th}(\bq;k)\right\}} +  \frac{\Gamma(\bq)}{\sqrt{\sigma}}{\bf1}_{\left\{\sigma\geq\sigma_{\rm th}(\bq;k)\right\}}.  
\end{align}
Thus, in view of \eqref{eq:Phi}, the buyer’s utility for purchasing the query $\bm{Q}=(\bm{q},\sigma)\in\R^n\times(0,\infty)$ is given by 
\begin{align}\label{eq:Phi22}
\Phi({\bm q},\sigma;k)&=v({\bm q},\sigma)-\pi({\bm q},\sigma;k)=\frac{A(\bm q)}{\sqrt{\sigma}}-\frac{kf(\bq)^{2}}{\sigma}{\bf1}_{\left\{\sigma\leq\sigma_{\rm th}(\bq;k)\right\}} -  \frac{\Gamma(\bq)}{\sqrt{\sigma}}{\bf1}_{\left\{\sigma\geq\sigma_{\rm th}(\bq;k)\right\}}\nonumber\\
&=\underbrace{\left(\frac{A(\bm q)}{\sqrt{\sigma}}-\frac{kf(\bq)^{2}}{\sigma}\right)}_{=:\Phi_1(\bqs;k)}{\bf1}_{\left\{\sigma\leq\sigma_{\rm th}(\bq;k)\right\}}+\underbrace{\frac{A(\bq)-\Gamma(\bq)}{\sqrt{\sigma}}}_{=:\Phi_2(\bqs;k)}{\bf1}_{\left\{\sigma\geq\sigma_{\rm th}(\bq;k)\right\}}
\end{align}
By using the first-order condition w.r.t. $\sigma$, it holds that
\begin{align}\label{eq:sigma1kstar0}
    \frac{\partial\Phi_1(\bqs;k)}{\partial\sigma}=0 \implies \text{the critical point }\sigma_1^*(\bq;k)=4\left(\frac{k}{A(\bq)}\right)^2f(\bq)^4.
\end{align}
Moreover, we have $\frac{\partial\Phi_1(\bqs;k)}{\partial\sigma}\geq0$ if $\sigma\leq\sigma_1^*(\bq;k)$, while $\frac{\partial\Phi_1(\bqs;k)}{\partial\sigma}\leq0$ if $\sigma\geq\sigma_1^*(\bq;k)$. Then, we consider the following situations under the case where $A(\bq)\geq\Gamma(\bq)$. In fact, $[\sigma_{\rm th}(\bq;k),\infty)\ni\sigma\to\Phi_2(\bqs;k)$ is strictly decreasing when $A(\bq)\geq\Gamma(\bq)$, we have $\Phi_2(\bqs;k)$ with $\sigma\geq\sigma_{\rm th}(\bq;k)$ has the maximizer at $\sigma=\max\{\sm(\bq),\sigma_{\rm th}(\bq;k)\}$ by considering the additional constraint $\sigma\geq\sm(\bq)$.
\begin{itemize}
    \item[(I)] when $\sigma_1^*(\bq,k)\geq \sigma_{\rm th}(\bq;k)$ (equivalently $A(\bq)\leq2\Gamma(\bq)$), we have $\sigma\to\Phi_1(\bqs;k)$ is strictly increasing as $\sigma\in(0,\sigma_{\rm th}(\bq;k))$, and hence $\sigma\to\Phi(\bqs;k)$ is strictly increasing as $\sigma\in(0,\sigma_{\rm th}(\bq;k)]$; while it is strictly decreasing as $\sigma\in[\sigma_{\rm th}(\bq;k),\infty)$. Thus, by the continuity of $\sigma\to\Phi(\bqs;k)$, we have $\Phi(\bqs;k)$ for $\sigma\geq\sm(\bq)$ has the maximizer at $\sigma^*(\bq;k)=\max\{\sm(\bq),\sigma_{\rm th}(\bq;k)\}$ by considering the additional constraint $\sigma\geq\sm(\bq)$ (c.f. Figure \ref{fig:placeholder2}).
    
    \begin{figure}[H]
        \centering
        \includegraphics[width=0.5\linewidth]{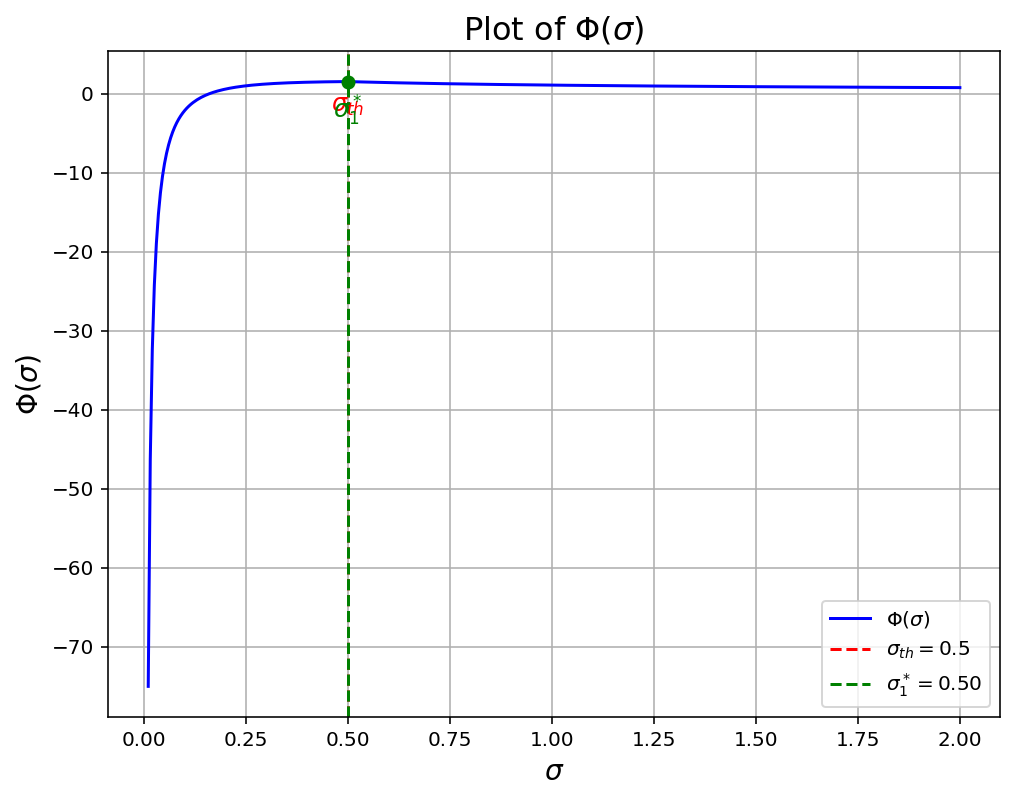}
        \caption{The utility $\sigma\to\Phi(\bqs;k)$ in the case where $\Gamma(\bq)\leq A(\bq)\leq 2\Gamma(\bq)$ ($A(\bq)=2.5$ and $\Gamma(\bq)=\sqrt{2}$).}
        \label{fig:placeholder2}
    \end{figure}

\item[(II)] when $\sigma_1^*(\bq,k)\leq\sigma_{\rm th}(\bq;k)$ (equivalently $A(\bq)\geq 2\Gamma(\bq)$), we have $\sigma\to\Phi_1(\bqs;k)$ is strictly increasing as $\sigma\in(0,\sigma_1^*(\bq;k)]$, and hence by the continuity of $\sigma\to\Phi(\bqs;k)$, $\sigma\to\Phi(\bqs;k)$ is strictly increasing as $\sigma\in(0,\sigma_1^*(\bq;k)]$; while it is strictly decreasing as $\sigma\in[\sigma_1^*(\bq;k),\infty)$. Thus, by the continuity of $\sigma\to\Phi(\bqs;k)$, we have $\Phi(\bqs;k)$ for $\sigma\geq\sm(\bq)$ has the maximizer at $\sigma^*(\bq,k)=\max\{\sm(\bq),\sigma^*_1(\bq;k)\}$ by considering the additional constraint $\sigma\geq\sm(\bq)$ (c.f. Figure \ref{fig:placeholder3}).

\begin{figure}[H]
        \centering
        \includegraphics[width=0.5\linewidth]{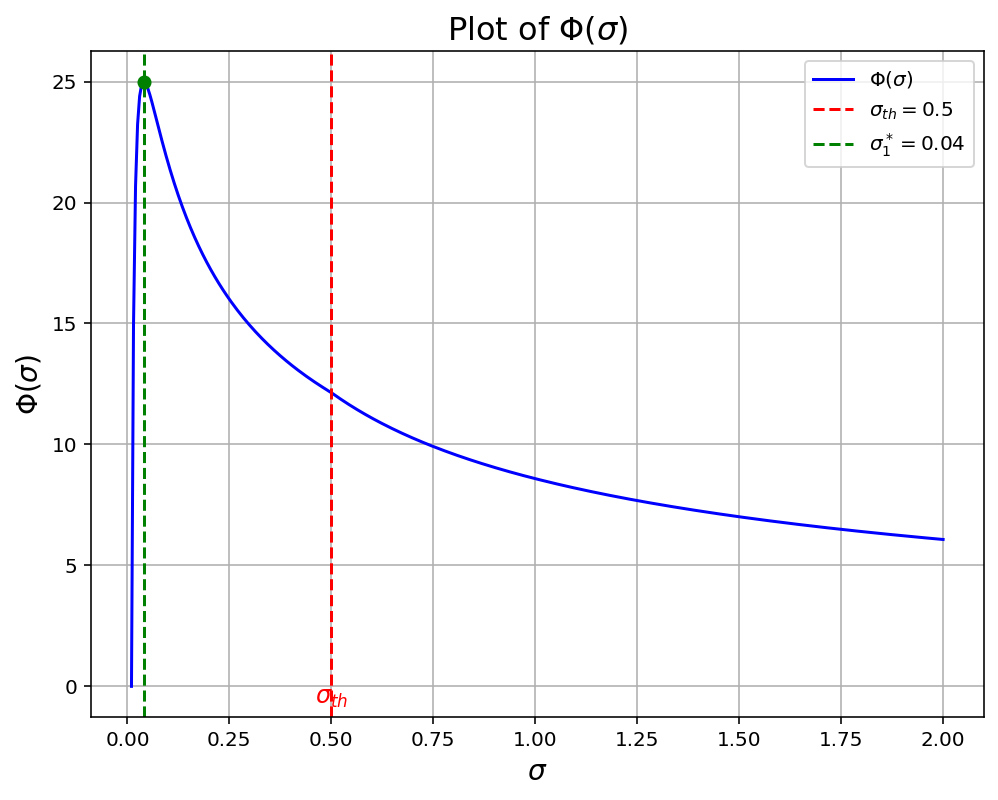}
        \caption{The utility $\sigma\to\Phi(\bqs;k)$ in the case where $A(\bq)\geq 2\Gamma(\bq)$ ($A(\bq)=10$ and $\Gamma(\bq)=\sqrt{2}$).}
        \label{fig:placeholder3}
    \end{figure}
\end{itemize}

\begin{figure}[H]
    \centering
    \includegraphics[width=0.5\linewidth]{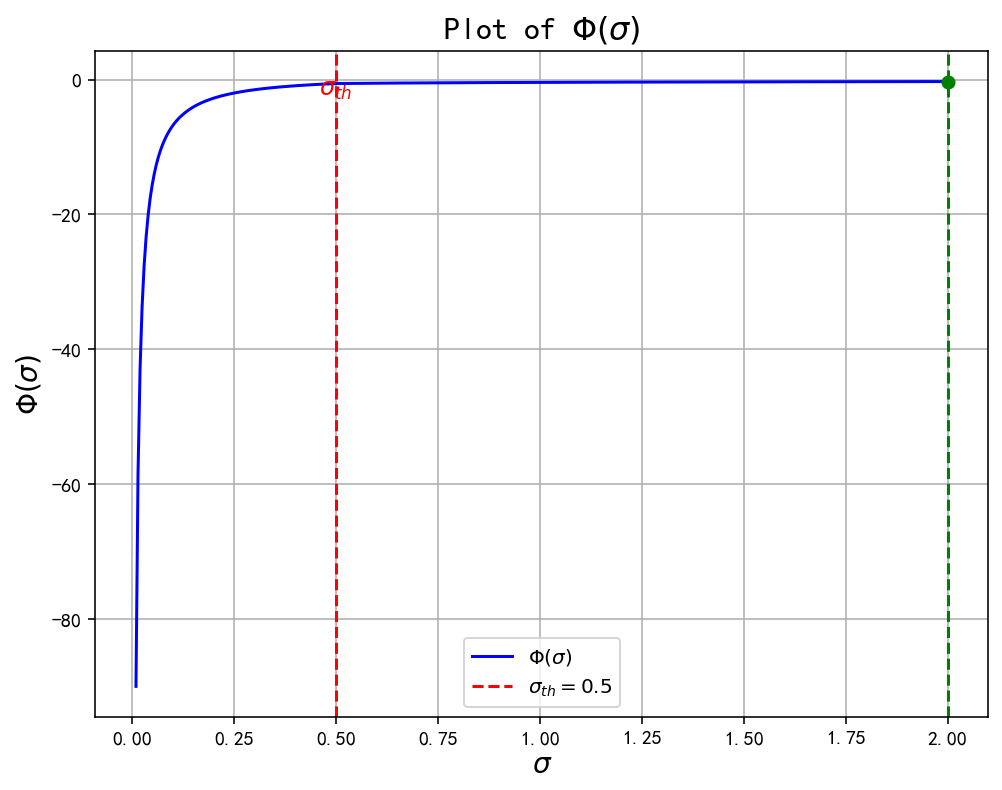}
    \caption{The utility $\sigma\to\Phi(\bqs;k)$ in the case where $A(\bq)\leq \Gamma(\bq)$ ($A(\bq)=1$ and $\Gamma(\bq)=\sqrt{2}$).}
    \label{fig:placeholder4}
\end{figure}

For the case $A(\bq)\leq\Gamma(\bq)$, we have  $\sigma\to\Phi_2(\bqs;k)$ is strictly increasing as $\sigma\in(0,\sigma_{\rm th}(\bq;k)]$ and $\sigma\to\Phi_2(\bqs;k)$ is also strictly increasing as $\sigma\in[\sigma_{\rm th}(\bq;k),\infty)$. Therefore, using the continuity of $\sigma\to\Phi(\bqs;k)$, it follows that  $\sigma\to\Phi(\bqs;k)$ is strictly increasing as $\sigma>0$. The overall trend of $\sigma\to\Phi(\bqs;k)$ is displayed in Figure \ref{fig:placeholder4}. Hence, the maximizer of $\Phi(\bqs;k)$ for $\sigma\geq\sm(\bq)$ in this case is given by $\sigma^*(\bq,k)=+\infty$.

Given the result $\sigma^*(\bq;k)$ of {\bf Stage II} provided above, we solve {\bf Stage I}. In view of \eqref{eq:mm_utility}, the utility function of market maker at $\sigma=\sigma^*(\bq;k)$ is given by \eqref{eq:mm_utility}, i.e., 
\begin{align}\label{eq:Psistar}
\Psi^*(\bq;k):=\Psi(\bq,\sigma^*(\bq;k),k)=\left(\frac{kf(\bm{q})^{2}}{\sigma^*(\bq,k)}-\frac{\Gamma(\bq)}{\sqrt{\sigma^*(\bq,k)}}\right)^+.
\end{align}
Our goal is to find the optimal pricing decision variable $k^*>0$ that maximizes the above utility function $\Psi^*(\bq;k)$. Building upon the results from \textbf{Stage II}, we discuss the following three cases:
\begin{itemize}

\item[(I)] when $\Gamma(\bq)\leq A(\bq)\leq2\Gamma(\bq)$, the buyer's optimal response obtained above is given by $\sigma^*(\bq;k)=\max\{\sm(\bq),\sigma_{\rm th}(\bq;k)\}$ with $\sigma_{\rm th}(\bq;k)$ being defined by \eqref{eq:prichange}. In this case, the utility function $\Psi^*(\bq;k)$ becomes from \eqref{eq:prichange} that
\begin{align}\label{eq:express-Psistar}
 \Psi^*(\bq;k)&=\left(\frac{kf(\bm{q})^{2}}{\sigma^*(\bq,k)}-\frac{\Gamma(\bq)}{\sqrt{\sigma^*(\bq,k)}}\right)^+\nonumber\\
 &=\begin{cases}
     \displaystyle \left(\frac{kf(\bm{q})^{2}}{\sigma_{\rm th}(\bq;k)}-\frac{\Gamma(\bq)}{\sqrt{\sigma_{\rm th}(\bq;k)}}\right)^+, & \sm(\bq)\leq\sigma_{\rm th}(\bq;k),\\[1.4em]
     \displaystyle~ \left(\frac{kf(\bm{q})^{2}}{\sm(\bq)}-\frac{\Gamma(\bq)}{\sqrt{\sm(\bq)}}\right)^+, & \sm(\bq)\geq \sigma_{\rm th}(\bq;k).
 \end{cases}
\end{align}
By using \eqref{eq:express-Psistar}, we have 
\begin{itemize}
    \item for the case $\sm(\bq)\leq\sigma_{\rm th}(\bq;k)$, we have the objective function $\Psi^*(\bq;k)=0$ for all $k>0$.  In other words, in this case, the market maker is indifferent to all $k>0$.
    \item for the case $\sm(\bq) \geq \sigma_{\rm th}(\bq;k)$, it follows from \eqref{eq:prichange} that $\frac{kf(\bm{q})^{2}}{\sm(\bq)}\leq\frac{\Gamma(\bq)}{\sqrt{\sm(\bq)}}$. This yields from \eqref{eq:express-Psistar} that 
    \begin{align*}
        \Psi^*(\bq;k)=\left(\frac{kf(\bm{q})^{2}}{\sm(\bq)}-\frac{\Gamma(\bq)}{\sqrt{\sm(\bq)}}\right)^+=0.
    \end{align*}
  Similarly to the above case, the market maker is also indifferent to all $k>0$.
\end{itemize}
To summarize, in the case of $\Gamma(\bq)\leq A(\bq)\leq 2\Gamma(\bq)$, the market maker is limited to achieving financial equilibrium, operating within a zero-profit region. Hence, any $k>0$ is an optimal decision variable for the market maker. Correspondingly, the buyer's optimal variance strategy is given by 
    \begin{align*}
      \sigma^*(\bq,k)=\left(\frac{k}{\Gamma(\bq)}\right)^2f(\bq)^4,\quad \forall k>0.  
    \end{align*}

\item[(II)] when $A(\bq)\geq 2 \Gamma(\bq)$, the buyer's optimal response is $\sigma^*(\bq,k)=\max\{\sm(\bq),\sigma_{1}^{*}(\bq;k)\}$ with  $\sigma_{1}^{*}(\bq;k)$ being defined by \eqref{eq:sigma1kstar0}. Then, in view of \eqref{eq:Psistar}, we arrive at
\begin{itemize}
    \item  for the case where $\sm(\bq) \leq \sigma_{1}^{*}(\bq;k)$ (this equivalently is $k\geq \bar{k}_1:= \frac{A(\bq)}{2f(\bq)^2}\sqrt{\sm(\bq)}$), we have the utility function of market maker becomes that
    \begin{align}\label{Psi3}
        \Psi^*(\bq;k)=\left(\frac{kf(\bm{q})^{2}}{\sigma_1^*(\bq;k)}-\frac{\Gamma(\bq)}{\sqrt{\sigma_1^*(\bq;k)}}\right)^+=\frac{A(\bq)}{4kf(\bq)^2}(A(\bq)-2\Gamma(\bq)).
    \end{align} 
As a consequence, the utility of market maker $k \to \Psi^*(\bq,\sigma;k)$ is strictly decreasing as $k \in [\bar{k}_1,\infty)$. 

\item for the case where $\sm(\bq) \geq \sigma_{1}^{*}(\bq;k)$ (this equivalently is $k\leq \bar{k}_1:= \frac{A(\bq)}{2f(\bq)^2}\sqrt{\sm(\bq)}$), we have the utility function of market maker becomes that
\begin{align}\label{eq:express-PsistarII}
 \Psi^*(\bq;k)&=\left(\frac{kf(\bm{q})^{2}}{\sigma^*(\bq,k)}-\frac{\Gamma(\bq)}{\sqrt{\sigma^*(\bq,k)}}\right)^+=\begin{cases}
     \displaystyle k\frac{f(\bm{q})^{2}}{\sm(\bq)}-\frac{\Gamma(\bq)}{\sqrt{\sm(\bq)}}, & k_1 \leq k\leq \bar{k}_1,\\[1.4em]
\displaystyle ~~~~~~~~~~0,& k\leq k_1,
 \end{cases}
\end{align}
where $k_1:=\frac{\Gamma(\bq)}{f(\bq)^2}\sqrt{\sm(\bq)}$ (since $A(\bq)\geq2\Gamma(\bq)$, we have $k_1\leq \bar{k}_1$). It corresponds exactly to the critical value at which the two branches of the pricing function $\pi(\bqs;k)$ are equal. Then, $k \to \Psi^*(\bq;k)$ is strictly increasing as $k \in (k_1,\bar{k}_1]$.
\end{itemize}
As summary, in the case with $A(\bq)\geq 2 \Gamma(\bq)$, the utility of market marker can be expressed by 
\begin{align}\label{eq:express-PsistarIIasasummary}
 \Psi^*(\bq;k)&=\left(\frac{kf(\bm{q})^{2}}{\sigma^*(\bq,k)}-\frac{\Gamma(\bq)}{\sqrt{\sigma^*(\bq,k)}}\right)^+
 =\begin{cases}
 \displaystyle \frac{A(\bq)}{4kf(\bq)^2}(A(\bq)-2\Gamma(\bq)), & k\geq\bar{k}_1,\\[1.4em]
     \displaystyle k\frac{f(\bm{q})^{2}}{\sm(\bq)}-\frac{\Gamma(\bq)}{\sqrt{\sm(\bq)}}, & k_1 \leq k\leq \bar{k}_1,\\[1.4em]
\displaystyle ~~~~~~~~~~~~~~~~0,& k\leq k_1.
 \end{cases}
\end{align}
Consequently, the optimal decision variable of market maker is given by 
\begin{align*}
    k^*(\bq)=\bar{k}=\frac{A(\bq)}{f(\bq)^2}\frac{\sqrt{\sm(\bq)}}{2}.
\end{align*}
Then, the corresponding optimal variance level is 
\begin{align*}
   \sigma^*(\bq)&=\max\{\sm(\bq),\sigma_{1}^{*}(\bq;k^*(\bq))\}=\max\left\{\sm(\bq),4\left(\frac{\bar{k}}{A(\bq)}\right)^2f(\bq)^4\right\}\nonumber\\
   &=\max\{\sm(\bq),\sm(\bq)\}=\sm(\bq). 
\end{align*}
The overall trend of the utility of market maker $k\to \Psi^*(\bq;k)$ is displayed in Figure \ref{Psi(k)} when $A(\bq)\geq 2 \Gamma(\bq)$.
\begin{figure}[H]
    \centering
\includegraphics[width=0.7\linewidth]{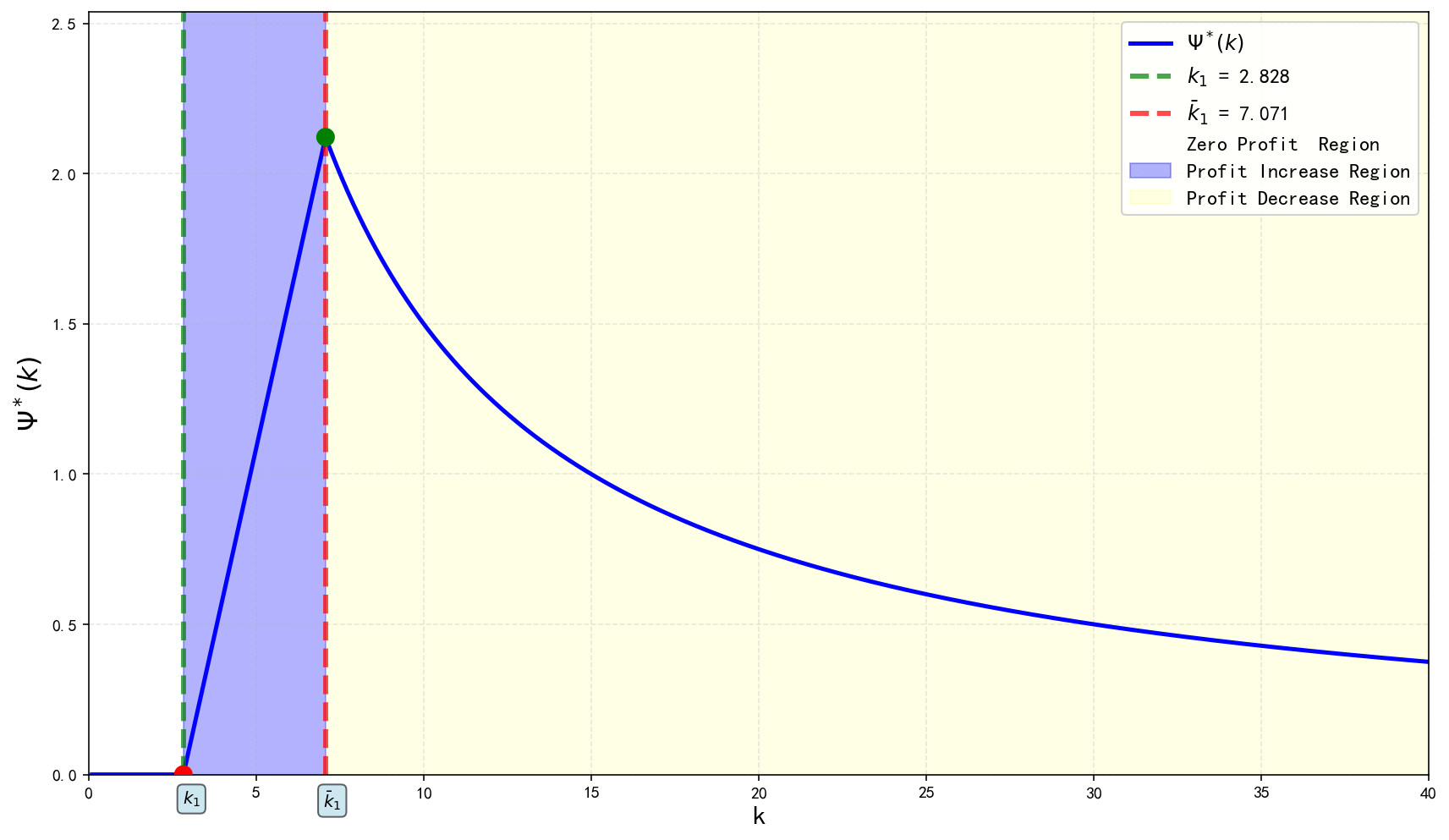}
    \caption{The utility of marker maker $k\to\Psi^*(\bq;k)$ when $A(\bq)\geq 2 \Gamma(\bq)$.}
    \label{Psi(k)}
\end{figure}
Figure \ref{Psi(k)} illustrates how the market maker's equilibrium profit $\Psi^*(\bq;k)$ varies with its pricing decision variable $k$ under a linear pricing function. The curve exhibits a single-peaked shape, clearly revealing the core trade-off faced by the market maker: setting the price too low ($k < k_1$) fails to cover the privacy compensation costs, resulting in zero profit; setting the price too high ($k > \bar{k}_1$) drives the data buyer to choose lower-precision data to avoid high prices, thereby geting profit. Therefore, the market maker's optimal strategy is to set $k^* = \bar{k}_1$, achieving profit maximization at the peak point. This figure visually verifies the conclusion of Theorem 1 that when the query value intensity is sufficiently high ($A(\bq) \geq 2\Gamma(\bq)$), the market maker can achieve an optimal balance between incentivizing the buyer to maintain high data precision and maximizing its own profit through precise pricing.

\item[(III)] The last case is $A(\bq)\leq\Gamma(\bq)$. In this case, the optimal variance $\sigma^*(\bq;k)=+\infty$ for fixed pricing decision variable $k>0$. Then, it follows form \eqref{eq:Psistar} that the utility of market marker $\Psi^*(\bq,k)= 0$ for all $k>0$. This concludes that any $k>0$ is the optimal decision variable of the market maker.  Thus, we complete the proof of the theorem. \hfill$\Box$
\end{itemize}

\section{Nonlinear Power Pricing Function}\label{sec:nonlineapricingfcn}

Section \ref{sec:model} considers the {\it balanced} pricing function $\pi(\bq;k)$ given by \eqref{pricing function}, in which, the pricing decision variable $k$ enters the original pricing function $\pi_0$ linearly (see \eqref{eq:originalpricingfcn}). In this section, we study the nonlinear pricing function w.r.t. the pricing decision variable $k$ in the following power nonlinearly form, for a query $\bQ=(\bqs)\in\R^n\times(0,\infty)$ and a pricing decision variable $k>0$,
\begin{align}\label{eq:pigf}
    \pi(\bqs;k) =\max \left\{ g(\pi_0(\bqs;k)),\sum_{i=1}^{n}\mu_{i}(\bq,\sigma)\right\},\quad \pi_0(\bqs;k)=k\frac{f(\bq)^{2}}{\sigma},
\end{align}
where the original pricing function $\pi_0$ depends on a function $f:\R^n\to\R$ and $g(x)=x^{p}$ with $p\in(\frac{1}{2},1]$ is a power utility function (c.f. \cite{Phelps24}). When the exponent $p=1$, the pricing function \eqref{eq:pigf} reduces to the one given by \eqref{pricing function} considered in the previous section.

We next assume that the function $f:\R^n\to\R$ is a {\it semi-norm} which can guarantee the arbitrage-freeness  of the original pricing function $\pi_0$ (c.f. Theorem 3.11 in \cite{Li2014}). 
The role of the function 
$g$ is to adjust the  pricing scheme by introducing a nonlinearity, which can better capture the complexities of market dynamics and strategic interactions between the market maker and the buyer. Note that $g$ is strictly increasing and concave on $(0,\infty)$. Namely, $g$ behaves as a classical power utility function in the sense that it is increasing in the effective precision level (\citet{Phelps24}), so that a reduction in the variance $\sigma$ leads to a higher precision level, while its concavity reflects the standard law of diminishing marginal utility with respect to information precision. The exponent $1-p$ refers to as the risk-aversion level of the power utility function as a CRRA (constant relative risk aversion) utility function.

We first prove that the nonlinear pricing function \eqref{eq:pigf} is arbitrage-free, which is provided in the following lemma:
\begin{lemma} \label{lem2}
Let $k>0$ be fixed. Then, for the query $\textbf{Q}=(\bq,\sigma)\in\mathbb{R}^{n}\times (0,\infty)$, the pricing function $\pi(\bq,\sigma;k)$ given by \eqref{eq:pigf} is arbitrage-free.
\end{lemma}

{\it Proof.}\quad We first show that the (nonnegative) concave function $g$ is sub-additive. In fact, it follows from the concavity of $g$ that,  for any $x,y>0$ and $\lambda\in(0,1)$, 
\begin{align}\label{eq:ine00}
    g(\lambda(x+y))= g\left(\lambda(x+y)+(1-\lambda)0\right)\geq \lambda g(x+y)+(1-\lambda) g(0)=\lambda g(x+y).
\end{align}
We next take $\lambda=\frac{x}{x+y}\in (0,1)$ and $\lambda=\frac{y}{x+y} \in (0,1)$, respectively. It follows from \eqref{eq:ine00} that, for any $x,y>0$, 
\begin{align*}
g(x) \geq \frac{x}{x+y} g(x+y),\quad g(y)\geq \frac{y}{x+y} g(x+y).
 \end{align*}
By adding the above two inequalities, we obtain $g(x+y)\leq g(x)+g(y)$ for any $x,y>0$. This shows that $(0,\infty)\ni x\to g(x)$ is sub-additive.

Next, we  verify that the pricing function 
$\pi_1(\bQ;k):=g(\pi_0(\bQ;k))$ is arbitrage-free. Following the proof of  Lemma \ref{lem:arbitrage-free}, we have the original pricing function $\pi_0(\bQ;k)$ is arbitrage-free. In other words, for any query $\bQ=(\bqs) \in \mathbb{R}^{n} \times (0, \infty)$ and linear answerable query sequence $\{\bQ_{j}=(\bq_{j},\sigma_{j})\}_{j=1}^{m}$, we have $\pi_{0}(\bQ;k)\leq \sum_{j=1}^{m}\pi_{0}(\bQ_{j};k)$. Then, by applying the non-decreasing property and the sub-addition of $g$, it follows that
\begin{align*}
    \pi_1(\bQ;k)=g\left(\pi_{0}(\bQ;k)\right) & \leq g\left(\sum_{j=1}^{m}\pi_{0}(\bQ_{j};k)\right)
    \leq \sum_{j=1}^{m}g\left(\pi_{0}(\bQ_{j};k)\right)=\sum_{j=1}^{m}\pi_1(\bQ_j;k).
\end{align*}
This yields that the pricing function $\pi_1(\bQ;k)= g(\pi_0(\bQ;k))$  is arbitrage-free. Moreover, we can conclude the arbitrage-freeness of the pricing function $\pi(\bQ;k)=\max\{\pi_1(\bQ;k),\sum_{i=1}^{n}\mu_{i}(\bQ)\}$ by applying Corollary 17 in \cite{Li2014}).
\hfill$\Box$

Within this nonlinear pricing framework, the buyer's utility for purchasing the query $\bm{Q}=(\bm{q},\sigma)\in\R^n\times(0,\infty)$ becomes that
\begin{align}\label{eq:Phi2}
        \Phi({\bm q},\sigma;k):=v({\bm q},\sigma)-\pi({\bm q},\sigma;k)=\frac{A({\bm q})}{\sqrt{\sigma}}-\max\left\{g\left(\frac{kf^2(\bq)}{\sigma}\right),\sum_{i=1}^{n}\mu_{i}(\bq,\sigma)\right\}.
\end{align}
On the other hand, the market maker, as the intermediary between data owners and buyers, also adapts its utility function to the new pricing scheme. The market maker’s objective is still to maximize its utility, which is the difference between the revenue received from the buyer and the compensation paid to the data owners. However, with the modified pricing function, the market maker's utility now takes the following form:
\begin{align} \label{eq:mm_objective2}
\Psi(\bm{q},\sigma;k):=\pi(\bq,\sigma;k)-\sum_{i=1}^{n}\mu_{i}(\bq,\sigma)=\left(g\left(\frac{kf^2(\bq)}{\sigma}\right)-\sum_{i=1}^{n}\mu_{i}(\bq,\sigma)\right)^{+}.
\end{align}

Now, recall the Stackelberg game problem \eqref{eq:Stage1} and \eqref{eq:Stage2} formulated in Section \ref{Game frame}. In this section, we shall introduce the nonlinear pricing function $\pi(\bq, \sigma;k)$ given by \eqref{eq:pigf} in the game problem. Then, we have
the following two-stage Stackelberg game problem described as follows:
\begin{itemize}
\item {\bf Stage I}~({\it Leader--Market Maker}). The market maker commits to a mechanism characterized by a price function $\pi(\bqs;k)$ and a compensation rule $\{\mu_i(\bqs)\}_{i=1}^n$, where $k>0$ is the mechanism parameter. The market maker’s objective is to choose an optimal pricing parameter $k^*(\bqs)$ so to maximize its own utility $\Psi(\bqs;k)$ over $k\in(0,\infty)$ when the query $\bQ=(\bqs)$ is given:
\begin{align}\label{eq:Stage122}
k^*(\bqs):= \arg\!\max_{k>0}\Psi(\bqs;k)=\arg\!\max_{k>0}\left[g(\pi_0(\bqs;k))-\sum_{i=1}^{n}\mu_i(\bqs)\right].
\end{align}

\item {\bf Stage II}~({\it Follower--Buyer}). Given $k^*(\bqs)$ provided by \eqref{eq:Stage122}. The buyer observes the announced price function and selects an optimal precision level (noise variance) $\sigma^*(\bq)$ to maximize its own utility $\Phi(\bqs;k^*(\bqs))$ over $\sigma\geq\sm(\bq)$, i.e.,
\begin{align}\label{eq:Stage222}
\sigma^*(\bq):=\arg\!\!\!\!\!\!\!\max_{\sigma\geq\sm(\bq)}\Phi(\bqs;k^*(\bqs))=\arg\!\!\!\!\!\!\!\max_{\sigma\geq\sm(\bq)}
\left[V(\bqs)-g(\pi_0(\bqs;k^*(\bqs)))\right].
    \end{align} 
\end{itemize}
We call $(k^*(\bq),\sigma^*(\bq))=(k^*(\bq,\sigma^*(\bq)),\sigma^*(\bq))$ the equilibrium strategy of the two-stage  Stackelberg game \eqref{eq:Stage122} and \eqref{eq:Stage222}.

\begin{Theorem}\label{The 2}
The buyer's utility $\Phi(\bqs;k)$ and the market-maker's utility $\Psi(\bqs;k)$ are respectively given by \eqref{eq:Phi2} and \eqref{eq:mm_objective2}. Then, the Stackelberg equilibrium strategy under the nonlinear pricing framework $(k^*(\bq),\sigma^*(\bq))=(k^*(\bq,\sigma^*(\bq)),\sigma^*(\bq))$ is given as follows:
    \begin{itemize}
    \item[(i)] if  $A(\bm q)\geq 2p\Gamma(\bq)$, the market maker’s optimal decision variable and the buyer's variance strategy are given by
    \begin{align}\label{eq:nolineoptimalks}
     \left(k^*(\bq),\sigma^*(\bq)\right)=\left(\left(\frac{A(\bq)}{2pf(\bq)^{2p}}\right)^{\frac{1}{p}}\times(\sm(\bq))^{\frac{2p-1}{2p}},~\sm(\bq)\right).  
    \end{align}

    \item[(ii)] if $\Gamma(\bq)<A(\bm q)<2p\Gamma(\bq)$, the utility of market maker becomes that $\Psi(\bm q,\sigma;k) \equiv0$. This implies that any $k>0$ is an optimal decision variable for the market maker. The buyer's optimal variance strategy is given by 
    \begin{align*}
      \sigma^*(\bq,k)=\left(\frac{k^pf(\bq)^{2p}}{\Gamma(\bq)}\right)^{\frac{2}{2p-1}},\quad \forall k>0.  
    \end{align*}
    
    \item[(iii)] if $A(\bm q) \leq \Gamma(\bq)$,  the utility of the buyer $\Phi(\bm q,\sigma;k) \leq 0$ and $(0,\infty)\ni\sigma\to\Phi(\bqs;k)$ is strictly increasing, and hence the optimal variance is $\sigma^*(\bq)=+\infty$ and the optimal decision variable is any $k>0$. In this case, the buyer does not purchase any data.
\end{itemize}
\end{Theorem}

\begin{figure}[H]
    \centering
\includegraphics[width=0.7\linewidth]{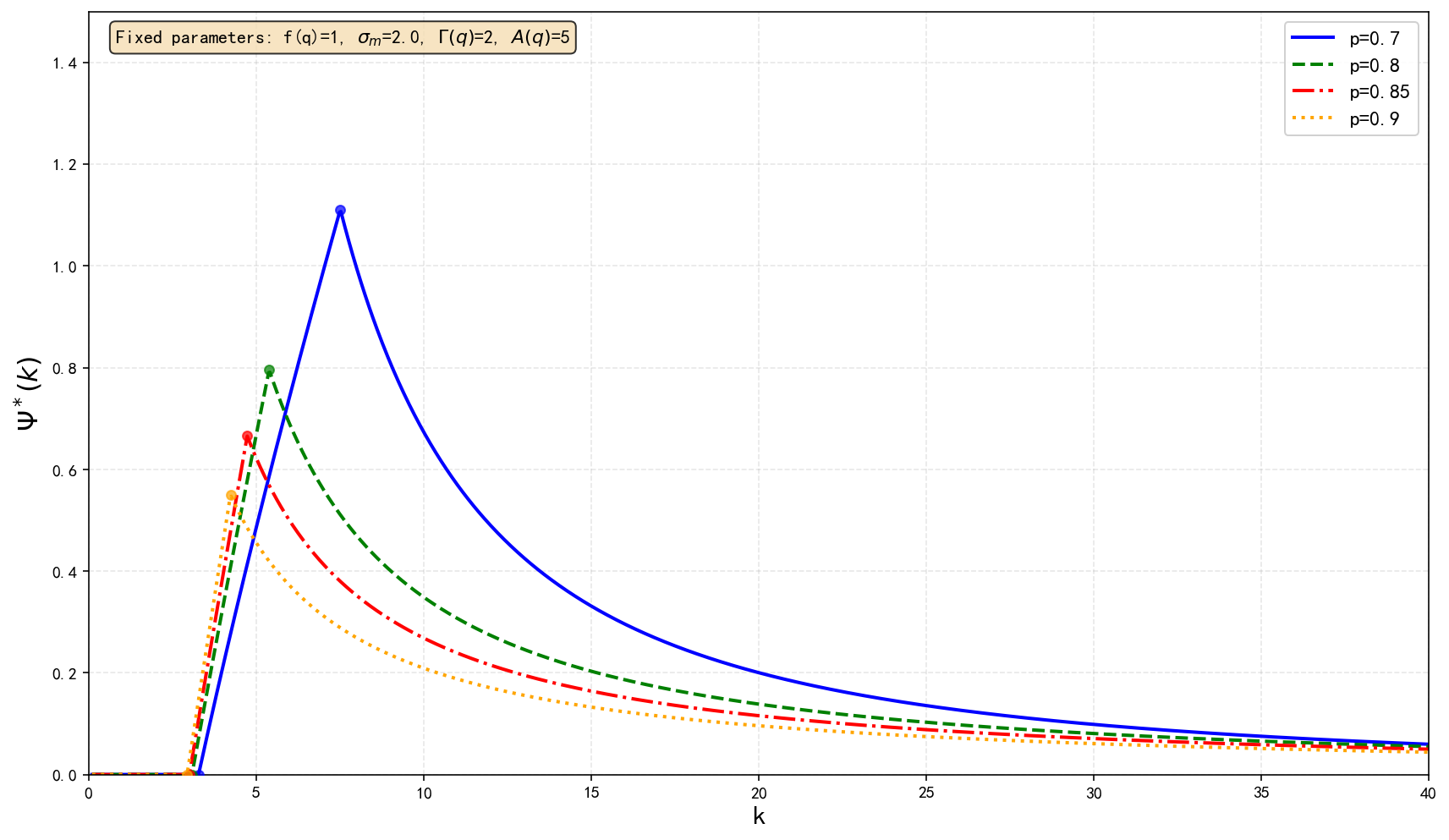}
    \caption{The market maker's profit function $\Psi^*(\bq;k)$ under varying degrees of risk aversion parameter $p$.}
    \label{pchange}
\end{figure}
Figure \ref{pchange} displays that, when $A(\bq)\geq2\Gamma(\bq)$, as the nonlinearity of the pricing function increases (i.e., as $p$ decreases), the peak profit of the market maker increases. From an economic standpoint, a smaller $p$ implies a lower elasticity of price with respect to data accuracy—the price increases relatively gradually as accuracy improves. This incentivizes the data buyer to choose queries with higher accuracy (lower noise variance $\sigma$), and higher-accuracy data inherently carries greater value for the buyer. Although the unit price increase per unit of accuracy is reduced, the higher accuracy level selected by the buyer still allows the market maker to achieve considerable profit. This finding indicates that in privacy-aware data markets, the market maker can optimize profit by adjusting the nonlinearity of the pricing function (the value of the exponent $p$): appropriately reducing the price elasticity with respect to accuracy (decreasing $p$) encourages the buyer to select data with higher-valued accuracy, leading to higher equilibrium profit.

\begin{figure}[ht]
    \centering
    \begin{subfigure}[b]{0.4\textwidth}
        \centering
\includegraphics[width=\textwidth]{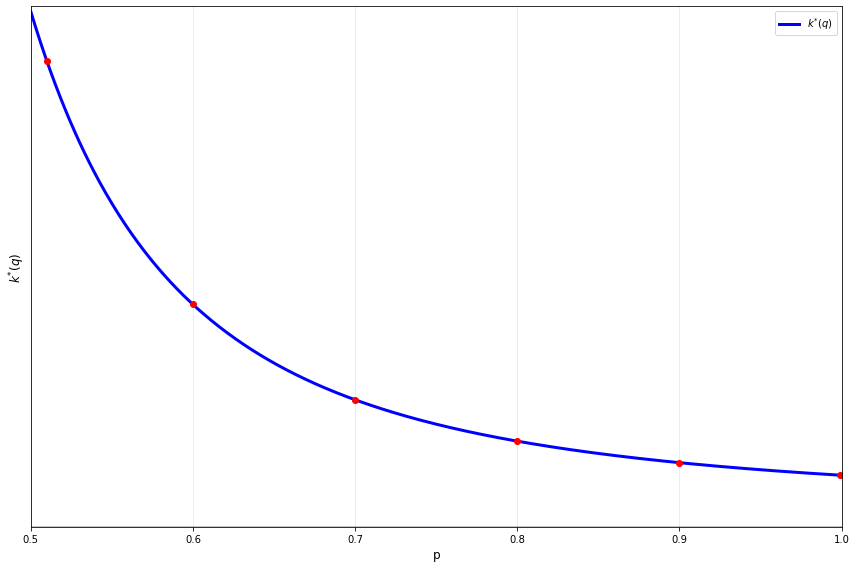}
        \caption{$p \mapsto k^{*}(\bq)$}
        \label{sub1}
    \end{subfigure}
    \hfill
    \begin{subfigure}[b]{0.4\textwidth}
        \centering
\includegraphics[width=\textwidth]{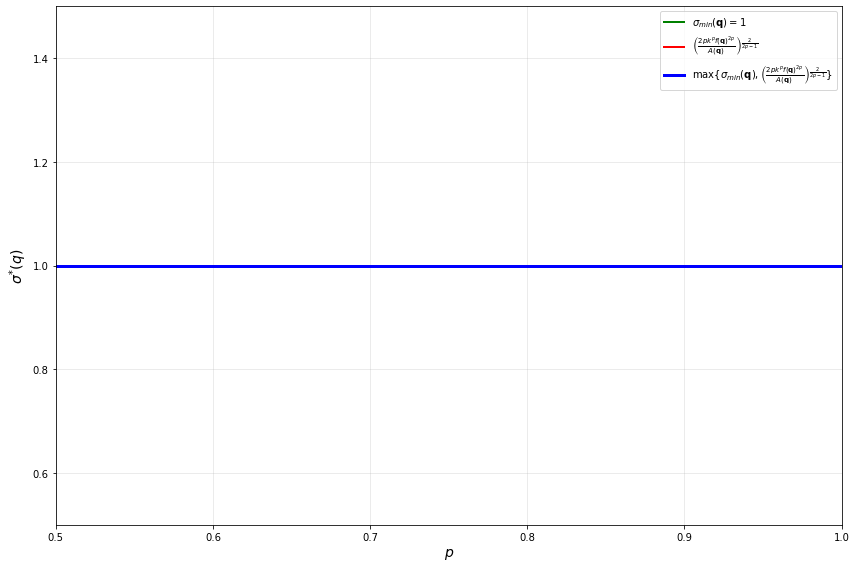}
        \caption{$p \mapsto \sigma^*(\bq)$}
        \label{fig:max_func}
    \end{subfigure}
    \caption{The Stackelberg equilibrium strategy $\left(k^*(\bq),\sigma^*(\bq)\right)=\left(\left(\frac{A(\bq)}{2pf(\bq)^{2p}}\right)^{\frac{1}{p}}(\sm(\bq))^{\frac{2p-1}{2p}},~\sm(\bq)\right)$,\\ which is given by \eqref{eq:nolineoptimalks} when $A(\bq)\geq 2p\Gamma(\bq)$ ($A(\bq)=5, f(q)=1, \sm(\bq)=1$).}
    \label{fig:functions}
\end{figure}

Figure \ref{fig:functions} shows that the optimal pricing decision $k^*(\bq)$ is decreasing with respect to the risk aversion parameter $p$. In fact, the parameter $p$ adjusts the buyer's marginal utility of precision. When $p$ is low, the buyer's utility saturates quickly with precision, making them highly sensitive to price increases. Anticipating this response, the market maker acting as the Stackelberg leader, optimally lowers $k^*(\bq)$ to prevent the buyer from retreating to a low-precision or no-trade outcome. Thus, $k^*(\bq)$ decreasing in $p$ is a direct consequence of strategic adaptation in a sequential game.

\begin{figure}[H]
    \centering
    \begin{subfigure}[b]{0.4\textwidth}
        \centering
\includegraphics[width=\textwidth]{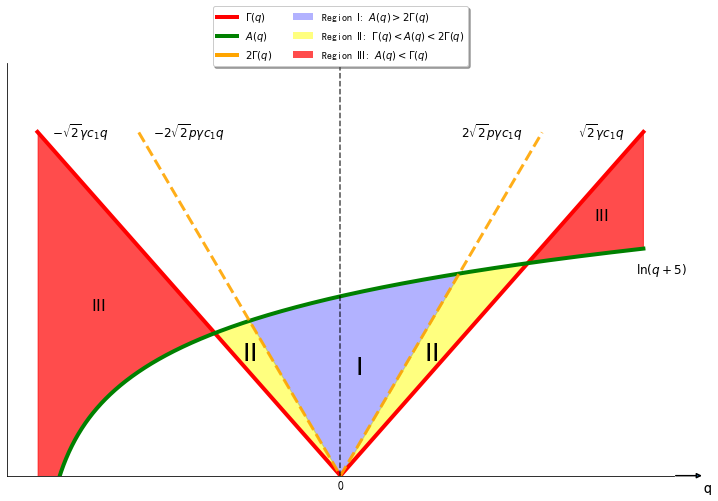}
        \caption{$p=0.75$}
        \label{sub3}
    \end{subfigure}
    \hfill
    \begin{subfigure}[b]{0.4\textwidth}
        \centering
\includegraphics[width=\textwidth]{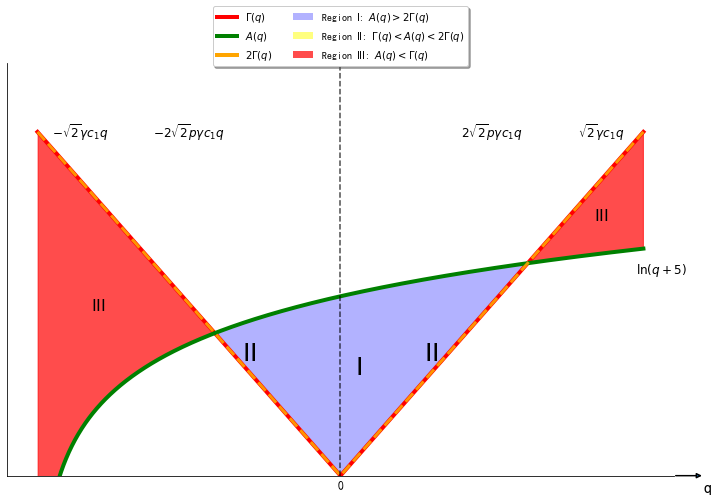}
        \caption{$p=0.5$}
        \label{sub4}
    \end{subfigure}
    \caption{The evolution of the trading region for different values of $p$.}
    \label{pregionchange}
\end{figure}
Figure \ref{pregionchange}  illustrates how three distinct trading regions evolve with changes in the exponent $p$. A smaller value of $p$ corresponds to a larger positive profit region for the market maker. This occurs because lower pricing attracts more buyers to opt for high-precision query results. As $p=\frac{1}{2}$, the zero-profit region gradually diminishes, compelling all buyers to either abstain from purchasing or acquire the high-precision query results.

We are now in a position to prove Theorem~\ref{The 2}.

\noindent{\it Proof of Theorem~\ref{The 2}.}\quad We first solve the above {\bf Stage II} when the decision variable $k>0$ is fixed. To this purpose, recall the pricing function $\pi(\bm{q},\sigma;k)$ given by \eqref{eq:pigf}. Then, we have
\begin{align}\label{eq:nolinepricingfcnpi}
\pi(\bm{q},\sigma;k)&=\max\left\{g\left(\frac{kf(\bm q)^{2}}{\sigma}\right),~\frac{\Gamma(\bq)}{\sqrt{\sigma}}\right\}\nonumber\\
&= g\left(\frac{kf(\bm q)^{2}}{\sigma}\right){\bf1}_{\left\{g\left(\frac{kf(\bm q)^{2}} {\sigma}\right)\geq\frac{\Gamma(\bq)}{\sqrt{\sigma}}\right\}} +  \frac{\Gamma(\bq)}{\sqrt{\sigma}}{\bf1}_{\left\{g\left(\frac{kf(\bm q)^{2}}{\sigma}\right)\leq\frac{\Gamma(\bq)}{\sqrt{\sigma}}\right\}}.  
\end{align}
Note that, for any $\sigma>0$, we have
\begin{align}\label{eq:nolineprichange2}
\left(\frac{kf(\bm q)^{2}}{\sigma}\right)^{p}\geq\frac{\Gamma(\bq)}{\sqrt{\sigma}}\iff \sigma \leq \tilde{\sigma}_{\rm th}(\bq;k):=\left(\frac{k^pf(\bq)^{2p}}{\Gamma(\bq)}\right)^{\frac{2}{2p-1}}.    
\end{align}
Consequently, it follows from \eqref{eq:nolinepricingfcnpi} that
\begin{align}\label{eq:pricingfcnpi2}
\pi(\bm{q},\sigma;k)&=\max\left\{\left(\frac{kf(\bm q)^{2}}{\sigma}\right)^{p},\ \frac{\Gamma(\bq)}{\sqrt{\sigma}}\right\}= \left(\frac{kf(\bm q)^{2}}{\sigma}\right)^{p}{\bf1}_{\left\{\sigma\leq\tilde{\sigma}_{\rm th}(\bq;k)\right\}} +  \frac{\Gamma(\bq)}{\sqrt{\sigma}}{\bf1}_{\left\{\sigma\geq\tilde{\sigma}_{\rm th}(\bq;k)\right\}}.  
\end{align}
Thus, in view of \eqref{eq:Phi2}, the buyer’s utility for purchasing the query $\bm{Q}=(\bm{q},\sigma)\in\R^n\times(0,\infty)$ is given by 
\begin{align}\label{eq:nolinePhi22}
\Phi({\bm q},\sigma;k)&=v({\bm q},\sigma)-\pi({\bm q},\sigma;k)=\frac{A(\bm q)}{\sqrt{\sigma}}-\left(\frac{kf(\bm q)^{2}}{\sigma}\right)^{p}{\bf1}_{\left\{\sigma\leq\tilde{\sigma}_{\rm th}(\bq;k)\right\}} -  \frac{\Gamma(\bq)}{\sqrt{\sigma}}{\bf1}_{\left\{\sigma\geq\tilde{\sigma}_{\rm th}(\bq;k)\right\}}\nonumber\\
&=\underbrace{\left(\frac{A(\bm q)}{\sqrt{\sigma}}-\left(\frac{kf(\bm q)^{2}}{\sigma}\right)^{p}\right)}_{=:\Phi_3(\bqs;k)}{\bf1}_{\left\{\sigma\leq\tilde{\sigma}_{\rm th}(\bq;k)\right\}}+\underbrace{\frac{A(\bq)-\Gamma(\bq)}{\sqrt{\sigma}}}_{=:\Phi_4(\bqs;k)}{\bf1}_{\left\{\sigma\geq\tilde{\sigma}_{\rm th}(\bq;k)\right\}}.
\end{align}
By using the first-order condition w.r.t. $\sigma$, it holds that
\begin{align}\label{eq:nolinesigma1kstar0}
    \frac{\partial\Phi_3(\bqs;k)}{\partial\sigma}=0 \implies \text{the critical point }\sigma_2^*(\bq;k)=\left(\frac{2pk^pf(\bq)^{2p}}{A(\bq)}\right)^{\frac{2}{2p-1}}.
\end{align}
Moreover, since $p>\frac{1}{2}$, we have $\frac{\partial\Phi_3(\bqs;k)}{\partial\sigma}\geq0$ if $\sigma\leq\sigma_2^*(\bq;k)$, while $\frac{\partial\Phi_3(\bqs;k)}{\partial\sigma}\leq0$ if $\sigma\geq\sigma_2^*(\bq;k)$. Then, we consider the following situations under the case where $A(\bq)\geq\Gamma(\bq)$. In fact, $[\tilde{\sigma}_{\rm th}(\bq;k),\infty)\ni\sigma\to\Phi_4(\bqs;k)$ is strictly decreasing when $A(\bq)\geq\Gamma(\bq)$, we have $\Phi_4(\bqs;k)$ with $\sigma\geq\tilde{\sigma}_{\rm th}(\bq;k)$ has the maximizer at $\sigma=\max\{\sm(\bq),\tilde{\sigma}_{\rm th}(\bq;k)\}$ by considering the additional constraint $\sigma\geq\sm(\bq)$.
\begin{itemize}
    \item[(I)] when $\sigma_2^*(\bq,k)\geq \tilde{\sigma}_{\rm th}(\bq;k)$ (equivalently $A(\bq)\leq2p\Gamma(\bq)$), we have $\sigma\to\Phi_3(\bqs;k)$ is strictly increasing as $\sigma\in(0,\tilde{\sigma}_{\rm th}(\bq;k))$, and hence $\sigma\to\Phi(\bqs;k)$ is strictly increasing as $\sigma\in(0,\tilde{\sigma}_{\rm th}(\bq;k)]$; while it is strictly decreasing as $\sigma\in[\tilde{\sigma}_{\rm th}(\bq;k),\infty)$. Thus, by the continuity of $\sigma\to\Phi(\bqs;k)$, we have $\Phi(\bqs;k)$ for $\sigma\geq\sm(\bq)$ has the maximizer at $\sigma^*(\bq;k)=\max\{\sm(\bq),\tilde{\sigma}_{\rm th}(\bq;k)\}$ by considering the additional constraint $\sigma\geq\sm(\bq)$ (c.f. Figure \ref{fig:placeholder6}).
    \begin{figure}[H]
        \centering
\includegraphics[width=0.5\linewidth]{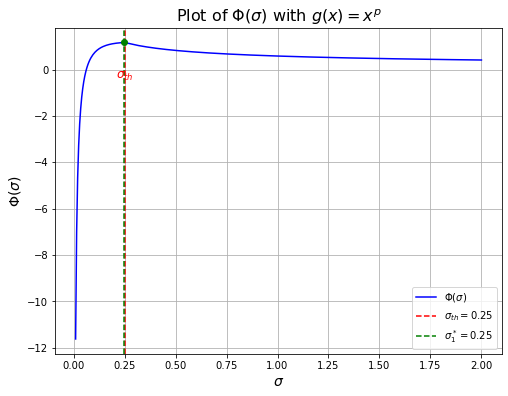}
        \caption{The utility $\sigma\to\Phi(\bqs;k)$ in the case where $\Gamma(\bq)\leq A(\bq)\leq 2p\Gamma(\bq)$\\ ($p=0.75, A(\bq)=2$ and $\Gamma(\bq)=\sqrt{2}$).}
    \label{fig:placeholder6}
    \end{figure}
\item[(II)] when $\sigma_2^*(\bq,k)\leq\tilde{\sigma}_{\rm{th}}(\bq;k)$ (equivalently $A(\bq)\geq 2p\Gamma(\bq)$), we have $\sigma\to\Phi_3(\bqs;k)$ is strictly increasing as $\sigma\in(0,\sigma_2^*(\bq;k)]$, and hence by the continuity of $\sigma\to\Phi(\bqs;k)$, $\sigma\to\Phi(\bqs;k)$ is strictly increasing as $\sigma\in(0,\sigma_2^*(\bq;k)]$; while it is strictly decreasing as $\sigma\in[\sigma_2^*(\bq;k),\infty)$. Thus, by the continuity of $\sigma\to\Phi(\bqs;k)$, we have $\Phi(\bqs;k)$ for $\sigma\geq\sm(\bq)$ has the maximizer at $\sigma^*(\bq,k)=\max\{\sm(\bq),\sigma^*_2(\bq;k)\}$ by considering the additional constraint $\sigma\geq\sm(\bq)$.
\begin{figure}[H]
        \centering
\includegraphics[width=0.5\linewidth]{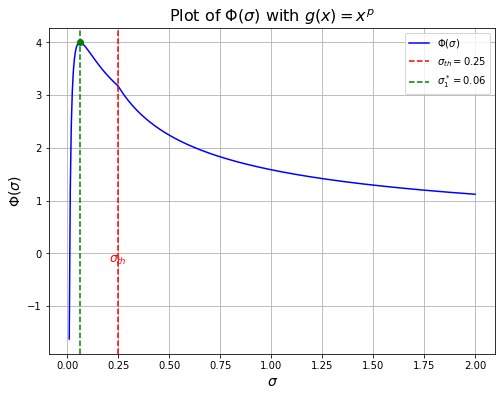}
        \caption{The utility $\sigma\to\Phi(\bqs;k)$ in the case where $ A(\bq)\geq 2p\Gamma(\bq)$\\ ($p=0.75, A(\bq)=5$ and $\Gamma(\bq)=\sqrt{2}$).}
    \label{fig:placeholder7}
    \end{figure}
\end{itemize}

Analogously to the case of linear pricing functions, for the case $A(\bq)\leq\Gamma(\bq)$, we have  $\sigma\to\Phi_4(\bqs;k)$ is strictly increasing as $\sigma\in(0,\tilde{\sigma}_{\rm th}(\bq;k)]$ and $\sigma\to\Phi_4(\bqs;k)$ is also strictly increasing as $\sigma\in[\tilde{\sigma}_{\rm th}(\bq;k),\infty)$. Therefore, using the continuity of $\sigma\to\Phi(\bqs;k)$, it follows that  $\sigma\to\Phi(\bqs;k)$ is strictly increasing as $\sigma>0$. Hence, the maximizer of $\Phi(\bqs;k)$ for $\sigma\geq\sm(\bq)$ in this case is given by $\sigma^*(\bq,k)=+\infty$. 
\begin{figure}[H]
        \centering
\includegraphics[width=0.5\linewidth]{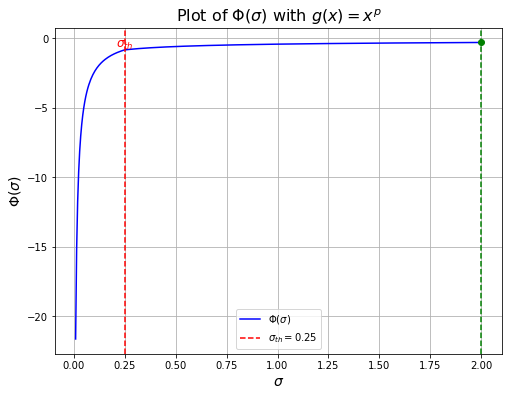}
        \caption{The utility $\sigma\to\Phi(\bqs;k)$ in the case where $A(\bq)\leq \Gamma(\bq)$ ($p=0.75, A(\bq)=1$ and $\Gamma(\bq)=\sqrt{2}$).}
    \label{fig:placeholder6}
    \end{figure}

Given the result $\sigma^*(\bq;k)$ of {\bf Stage II} provided above, we solve {\bf Stage I}. In view of \eqref{eq:Phi2}, the utility function of market maker at $\sigma=\sigma^*(\bq;k)$ is given by \eqref{eq:Phi2}, i.e., 
\begin{align}\label{eq:Psistar2}
\Psi^*(\bq;k):=\Psi(\bq,\sigma^*(\bq;k),k)=\left(\left(\frac{kf(\bm{q})^{2}}{\sigma^*(\bq,k)}\right)^{p}-\frac{\Gamma(\bq)}{\sqrt{\sigma^*(\bq,k)}}\right)^+.
\end{align}
Based on the analysis in \textbf{Stage II}, now we consider the following three cases:
\begin{itemize}
    \item [(I)]  When $\Gamma(\bq)\leq A(\bq)\leq2p\Gamma(\bq)$ for any $p\in(\frac{1}{2},1]$, the buyer's optimal response obtained above is given by $\sigma^*(\bq;k)=\max\{\sm(\bq),\tilde{\sigma}_{\rm th}(\bq;k)\}$ with $\tilde{\sigma}_{\rm th}(\bq;k)$ being defined by \eqref{eq:nolineprichange2}. In this case, the utility function $\Psi^*(\bq;k)$ becomes from \eqref{eq:nolineprichange2} that
\begin{align}\label{eq:nolineexpress-Psistar}
 \Psi^*(\bq;k)&=\left(\left(\frac{kf(\bm{q})^{2}}{\sigma^*(\bq,k)}\right)^{p}-\frac{\Gamma(\bq)}{\sqrt{\tilde{\sigma}(\bq,k)}}\right)^+\nonumber\\
 &=\begin{cases}
     \displaystyle \left(\left(\frac{kf(\bm{q})^{2}}{\tilde{\sigma}_{\rm th}(\bq;k)}\right)^{p}-\frac{\Gamma(\bq)}{\sqrt{\tilde{\sigma}_{\rm th}(\bq;k)}}\right)^+, & \sm(\bq)\leq\tilde{\sigma}_{\rm th}(\bq;k),\\[1.4em]
     \displaystyle~~ \left(\left(\frac{kf(\bm{q})^{2}}{\sigma_{\rm min}(\bq)}\right)^{p}-\frac{\Gamma(\bq)}{\sqrt{\sm(\bq)}}\right)^+, & \sm(\bq)\geq \tilde{\sigma}_{\rm th}(\bq;k).
 \end{cases}
\end{align}
By using \eqref{eq:nolineexpress-Psistar}, we have 
\begin{itemize}
    \item for the case $\sm(\bq)\leq\tilde{\sigma}_{\rm th}(\bq;k)$, we have the objective function $\Psi^*(\bq;k)=0$ for all $k>0$.  In other words, in this case, the market maker is indifferent to all $k>0$.
    \item for the case $\sm(\bq) \geq \tilde{\sigma}_{\rm th}(\bq;k)$, it follows from \eqref{eq:nolineprichange2} that $\left(\frac{kf(\bm{q})^{2}}{\sm(\bq)}\right)^{p}\leq\frac{\Gamma(\bq)}{\sqrt{\sm(\bq)}}$. This yields from \eqref{eq:nolineexpress-Psistar} that 
    \begin{align*}
        \Psi^*(\bq;k)=\left(\left(\frac{kf(\bm{q})^{2}}{\sm(\bq)}\right)^p-\frac{\Gamma(\bq)}{\sqrt{\sm(\bq)}}\right)^+=0.
    \end{align*}
As in the preceding case, the market maker derives the same utility from any $k>0$.
\end{itemize}
Overall, when $\Gamma(\bq)\leq A(\bq)\leq 2p\Gamma(\bq)$, the market maker is constrained to a break-even position, operating within a zero-profit region. Hence, all $k>0$ are equally optimal from the market maker's perspective.
Correspondingly, the buyer's optimal variance strategy is given by 
    \begin{align*}
      \sigma^*(\bq,k)=\left(\frac{2pk^pf(\bq)^{2p}}{A(\bq)}\right)^{\frac{2}{2p-1}},\quad \forall k>0.  
    \end{align*}

\item[(II)] When $A(\bq)\geq2p\Gamma(\bq)$,  the buyer's optimal response is $\sigma^*(\bq,k)=\max\{\sm(\bq),\sigma_{2}^{*}(\bq;k)\}$ with  $\sigma_{2}^{*}(\bq;k)$ being defined by \eqref{eq:nolinesigma1kstar0}. Then, in view of \eqref{eq:mm_objective2}, we get that
\begin{itemize}
    \item  for the case with $\sm(\bq) \leq \sigma_{2}^{*}(\bq;k)$ which is equivalently $k\geq \bar{k}_2$, where 
    \begin{align}\label{eq:bark2}
     \bar{k}_2:= \left(\frac{A(\bq)}{2pf(\bq)^{2p}}\right)^{\frac{1}{p}}\times(\sm(\bq))^{\frac{2p-1}{2p}}, 
    \end{align}
we have the utility function of market maker becomes that
    \begin{align}\label{nolinePsi3}
        \Psi^*(\bq;k)&=\left(\left(\frac{kf(\bm{q})^{2}}{\sigma_2^*(\bq;k)}\right)^p-\frac{\Gamma(\bq)}{\sqrt{\sigma_2^*(\bq;k)}}\right)^+\nonumber\\
        &=\left(\frac{A(\bq)}{2pk^p f(\bq)^{2p}}\right)^{\frac{1}{2p-1}}\times
\left(\frac{A(\bq)}{2p}-\Gamma(\bq)\right).
    \end{align} 
As a consequence, the utility of market maker $k \to \Psi^*(\bq,\sigma;k)$ is strictly decreasing as $k \in [\bar{k}_2,\infty)$. 

\item for the case with $\sm(\bq) \geq \sigma_{2}^{*}(\bq;k)$ (this equivalently is $k\leq \bar{k}_2$), we have that the utility function of market maker becomes that
\begin{align}\label{eq:express-PsistarII}
 \Psi^*(\bq;k)&=\left(\left(\frac{kf(\bm{q})^{2}}{\sigma^*(\bq,k)}\right)^p-\frac{\Gamma(\bq)}{\sqrt{\sigma^*(\bq,k)}}\right)^+\nonumber\\
 &=\begin{cases}
     \displaystyle \left(k\frac{f(\bm{q})^{2}}{\sm(\bq)}\right)^{p}-\frac{\Gamma(\bq)}{\sqrt{\sm(\bq)}}, & k_2 \leq k\leq \bar{k}_2,\\[1.4em]
\displaystyle ~~~~~~~~~~~~~~~~~~0,& ~~~~k\leq k_2,
 \end{cases}
\end{align}
where the quantity $k_2$ is given by 
\begin{align}\label{eq:k2express}
k_2:=\left(\frac{\Gamma(\bq)}{f(\bq)^{2p}}\right)^{\frac{1}{p}}\times(\sm(\bq))^{\frac{2p-1}{2p}}.    
\end{align}
Since $A(\bq)\geq 2p\Gamma(\bq)$, we have $k_2\leq \bar{k}_2$. It corresponds exactly to the critical value at which the two branches of the pricing function $\pi(\bqs;k)$ are equal. Then, $k \to \Psi^*(\bq;k)$ is strictly increasing as $k \in (k_2,\bar{k}_2]$.
\end{itemize}

As summary, in the case with $A(\bq)\geq 2 p\Gamma(\bq)$, the utility of market marker can be expressed by 
\begin{align}\label{eq:nolineexpress-PsistarIIasasummary}
 \Psi^*(\bq;k)&=\left(\left(\frac{kf(\bm{q})^{2}}{\sigma^*(\bq,k)}\right)^p-\frac{\Gamma(\bq)}{\sqrt{\sigma^*(\bq,k)}}\right)^+\nonumber\\
 &=\begin{cases}
 \displaystyle \left(\frac{A(\bq)}{2pk^p f(\bq)^{2p}}\right)^{\frac{1}{2p-1}}\times
\left(\frac{A(\bq)}{2p}-\Gamma(\bq)\right), &~~~k\geq\bar{k}_2,\\[1.4em]
     \displaystyle~~~~~~\left(\frac{kf(\bm{q})^{2}}{\sm(\bq)}\right)^{p}-\frac{\Gamma(\bq)}{\sqrt{\sm(\bq)}}, & k_2 \leq k\leq \bar{k}_2,\\[1.4em]
\displaystyle ~~~~~~~~~~~~~~~~~~~~~~~~0,& ~~~~k\leq k_2.
 \end{cases}
\end{align}
Consequently, the optimal decision variable of market maker is given by 
\begin{align*}
    k^*(\bq)=\bar{k}_2= \left(\frac{A(\bq)}{2pf(\bq)^{2p}}\right)^{\frac{1}{p}}\times (\sm(\bq))^{\frac{2p-1}{2p}}.
\end{align*}
Then, the corresponding optimal variance level is given by
\begin{align*}
   \sigma^*(\bq)&=\max\{\sm(\bq),\sigma_{2}^{*}(\bq;k^*(\bq))\}=\max\left\{\sm(\bq),\left(\frac{2pk^pf(\bq)^{2p}}{A(\bq)}\right)^{\frac{2}{2p-1}}\right\}\nonumber\\
   &=\max\{\sm(\bq),\sm(\bq)\}=\sm(\bq). 
\end{align*}   
\end{itemize}
In the case where $A(\bq)\leq\Gamma(\bq)$, the optimal variance $\sigma^*(\bq;k)=+\infty$ for any market maker pricing decision variable $k>0$. Consequently, it follows from \eqref{eq:Psistar2} that the market maker's utility $\Psi^*(\bq,k)= 0$ for all $k>0$. This implies that any $k>0$ is an optimal decision variable for the market maker. Thus, we complete the proof of the theorem. \hfill$\Box$

\begin{remark}
We consider the power function $g(x)=x^p$ for the risk aversion parameter $p\in(\frac{1}{2},1]$. When $p>1$, the pricing function $\pi(\bqs;k)$ given by \eqref{eq:pigf} is no longer arbitrage-free. When $p=\frac{1}{2}$, our pricing framework shall degenerate to the trivial case: the buyer’s utility function is strictly decreasing in $\sigma\geq\sigma_{\rm min}(\bq)$, and hence the optimal choice of the variance is always $\sigma^*(\bq)=\sigma_{\min}(\bq)$. When $p<\frac{1}{2}$, the zero-profit region disappears; even for low-value-intensity queries, the buyer is forced to choose high-precision query results with relatively high prices.
\end{remark}

\section{Conclusion}\label{sec:conclu}

In this paper, we have developed a Stackelberg game-theoretic framework for pricing privacy-sensitive data queries, explicitly modeling the strategic interaction between a profit-maximizing market maker and a utility-maximizing data buyer under differential privacy constraints. Our model departs from traditional static or one-sided pricing schemes by capturing the sequential decision-making process inherent in data markets: the market maker  first commits to a pricing rule, which incorporates both revenue goals and mandatory privacy compensation costs, and the buyer  then optimally selects the query precision (noise level). We derived closed-form equilibrium solutions for both linear and nonlinear  pricing function, which clearly delineate three distinct market regimes—profitable trade, break-even, and no-trade—based on the relationship between the query's value intensity and the aggregate privacy cost threshold. These results provide a principled and analytically tractable foundation for designing incentive-compatible, arbitrage-free, and privacy-conscious data markets, highlighting how key parameters like the privacy cost coefficient and the nonlinearity of the pricing function critically shape market outcomes and efficiency.

We identify several promising directions for extending this work. First, a natural progression is to model dynamic data pricing, where data value and privacy budgets evolve over time. This could involve  the buyer adjust their strategies based on accumulated information or depleting privacy allowances. Second, introducing data sellers as active game players would enhance realism. Current work aggregates their compensation as a cost threshold; future models could endow sellers with strategic choices creating a multi-layer game that captures the full data supply chain's incentives. Third, exploring more complex, generalized nonlinear pricing functions could better align with intricate market realities. While this pursuit increases model expressiveness, it often comes at the cost of analytical tractability, potentially yielding non-closed-form solutions that necessitate computational or simulation-based methods for equilibrium analysis.

\end{document}